\documentclass[structabstract]{aa}

\usepackage{natbib}

\usepackage{balance}

\usepackage[ruled]{algorithm2e}
\usepackage{hyperref}
\usepackage{amsmath}
\usepackage{numprint}
\npthousandsep{,}\npthousandthpartsep{}\npdecimalsign{.}

\usepackage{float}

\usepackage{tikz}
\usetikzlibrary{shapes,arrows}
 
\usepackage{dblfloatfix}
\usepackage{fixltx2e}
\usepackage[font=scriptsize,labelfont=bf,up,textfont=rm,up]{caption}
\usepackage[font=scriptsize,labelfont=bf,textfont=rm]{subcaption}
\usepackage{xfrac}
 
\hypersetup{colorlinks=true,citecolor=blue}
 \usepackage{amsfonts,upgreek,bbm,mathrsfs}
\usepackage{bbm}

\usepackage{url}
\usepackage{amssymb}
\usepackage{amsfonts}
\usepackage[ruled]{algorithm2e}
\usepackage{graphicx,epsfig,graphics}
\usepackage{aas_macros}

\usepackage{verbatim}

\usepackage{footnote}

\usepackage[nodisplayskipstretch]{setspace}
\numberwithin{equation}{section}

\let\oldsqrt\sqrt
\def\sqrt{\mathpalette\DHLhksqrt}
\def\DHLhksqrt#1#2{%
\setbox0=\hbox{$#1\oldsqrt{#2\,}$}\dimen0=\ht0
\advance\dimen0-0.4\ht0
\setbox2=\hbox{\vrule height\ht0 depth -\dimen0}%
{\box0\lower0.4pt\box2}}
\newcommand{\norm}[1]{\left\|{#1}\right\|}
\newcommand{\st}{\mathrm{\quad s.t. \quad}}
\newcommand{\abs}[1]{\left|#1\right|}
\newcommand{\capt}[1]{\vspace{-16pt}\caption{#1}}
\newcommand{\sect}[1]{\S\,\ref{#1}}

\begin{document}

\label{firstpage}
\title{Darth Fader: Using wavelets to obtain accurate redshifts of spectra at very low signal-to-noise} 

\author{D. P. Machado\inst{1}
\thanks{\email{daniel.machado@cea.fr}}
\and A. Leonard\inst{1}
%\thanks{\email{adrienne.leonard@cea.fr}}
\and J.-L. Starck\inst{1}
%\thanks{\email{jstarck@cea.fr}}
\and F. B. Abdalla\inst{2}
%\thanks{\email{fba@star.ucl.ac.uk}}
\and S. Jouvel\inst{2,3}
}
\institute{CEA Saclay, IRFU, Service d'Astrophysique, Bt 709 - Orme des Merisiers, 91191 Gif-Sur-Yvette CEDEX, France. \and University College London, Department of Physics \& Astronomy, Kathleen Lonsdale Building, Gower Place, London, WC1E 6BT, United Kingdom. \and Institut de Ci\`{e}ncies de l'Espai (IEEC-CSIC), E-08193 Bellaterra (Barcelona), Spain.}

%\date{} % delete this line to display the current date

%%% BEGIN DOCUMENT

\abstract{
Accurate determination of the redshifts of galaxies comes from the identification of key lines in their spectra. Large sky surveys, and the sheer volume of data they produce,
 have made it necessary to tackle this identification problem in an automated and reliable fashion. Current methods attempt to do this with careful modelling of the spectral lines and the continua, or by employing a flux/magnitude or a signal-to-noise cut to the dataset in order to obtain reliable redshift estimates for the majority of galaxies in the sample.}
{In this paper, we present the DARTH FADER algorithm (\textbf{D}enoised and \textbf{A}utomatic \textbf{R}edshifts \textbf{Th}resholded with a \textbf{Fa}lse \textbf{De}tection \textbf{R}ate), which is a new wavelet-based method for estimating redshifts of galaxy spectra. Automated, simple, and largely empirical, we highlight how the Darth Fader algorithm performs in the very low signal-to-noise regime, and demonstrate its effectiveness at removing catastrophic failures from the catalogue of redshift estimates.}
{We present a new, nonparametric method for estimating and removing the continuum in noisy data that requires no \emph{a priori} information about the galaxy properties. This method employs wavelet filtering based on a tuneable false detection rate (FDR) threshold, which effectively removes the noise in the spectrum, and extracts features at different scales. After removal of the continuum, the galaxy spectra are then cross-correlated with the eigentemplates, and a standard $\chi^2$-minimisation used to determine the redshift of the spectrum. FDR filtering is applied to the spectra a second time to determine the number of spectral features in each galaxy spectrum, and those with fewer than six total features are removed from the catalogue as we are unlikely to obtain a reliable and correct estimate of the redshift of such spectra.}
{ Applying our wavelet-based cleaning algorithm on a simulated testing set, we can successfully build a clean catalogue including extremely low signal-to-noise data (SNR=2.0), for which we are able to obtain a 5.1\% catastrophic failure rate in the redshift estimates (compared with 34.5\% prior to cleaning). We also show that for a catalogue with uniformly mixed signal-to-noise ratios between 1.0 and 20.0, with realistic pixel-dependent noise, it is possible to obtain redshifts with a catastrophic failure rate of 3.3\% after cleaning (as compared to 22.7\% before cleaning). Whilst we do not test this algorithm exhaustively on real data, we present a proof of concept of the applicability of this method to real data, showing that the wavelet filtering techniques perform well when applied to some typical spectra from the Sloan Digital Sky Survey archive.}
{The Darth Fader algorithm provides a robust method for extracting spectral features from very noisy spectra: spectra for which a reliable redshift cannot be measured are automatically identified and removed from the input data set. The resulting clean catalogue, although restricted in number, gives an extremely low rate of catastrophic failures, even when the spectra have a very low SNR. For very large sky surveys, this technique may offer a significant boost in the number of faint galaxies with accurately determined redshifts.}


\titlerunning{Darth Fader: Using wavelets to obtain accurate spec-z at very low SNR}
%\authorrunning{}
\maketitle
%\clearpage

\section{Introduction}\label{sec:intro}

The simplest method for estimating the redshift of a galaxy spectrum is by visual inspection, however, large sky surveys are providing astronomers with increasingly large datasets. The sheer number of spectra being obtained in these surveys make it necessary to make use of automated algorithms to obtain accurate information, as well as sophisticated techniques for dealing with the presence of noise; something which is increasingly important for distant and dimmer sources.

Traditional methods for automated estimation of the redshifts of galaxy spectra have primarily been reliant upon template matching with cross-correlations \citep{Tonry:1979,Glazebrook:1998,Aihara:2011} or -- and sometimes in conjunction with -- the matching of spectral lines \citep{Kurtz:1998,Garilli:2010,Stoughton:2002}.

Spectral line matching methods involve the use of spectra with a high enough signal-to-noise ratio to detect at least one emission line above a predefined threshold. With multiple emission lines, it is a task of matching the respective rest frame wavelengths of lines such as H$_{\alpha}$ and the [O$\,$III] doublet, to their respective redshifted counterparts. When faced with just a single emission line, assuming it to be H$_{\alpha}$ or [O$\,$II] \numprint{3727}\,\textup{\AA} is a viable option for spectroscopic redshift determination of emission line galaxies \citep[since one of these is usually, but not always, the strongest feature;][]{LeFevre:1995,LeFevre:2005}. In such cases, degeneracies on the lines may potentially be resolved with the inclusion of photometric data. This type of approach is used in the SDSS Early Data Release \citep{Stoughton:2002}.

Template matching methods for redshift estimation involve a `test set' -- a catalogue of galaxy spectra with unknown redshifts -- being matched to a set of template spectra initially at zero redshift (`a template set'). The template spectra are matched to the data, typically using a $\chi^2$ test or maximum likelihood estimator to determine the shift in wavelength between the template and the galaxy spectrum, and hence the redshift of that galaxy. Templates may come from simulated or semi-empirical spectra based on sophisticated modelling, local galaxy spectra whose redshifts are small and precisely known, or from a subset of high signal-to-noise spectra within the survey itself with redshifts that can be confidently identified.

Cross-correlation methods such as that described by \citet{Glazebrook:1998} use a discrete Fourier transform to correlate a template spectrum with a galaxy spectrum allowing the shift of the template spectrum (and thus redshift) to become a free parameter. Cross-correlation methods are convenient because they can be computed as a simple multiplication in Fourier space between the template and galaxy spectra, resulting in easier and faster computation than performing the same procedure in real space. These kinds of cross-correlation methods, however, require the spectra to be free of continuum in order to correctly correlate line features, with the presence of lines being a stronger constraint for redshift estimation than the shape of the continuum. Failure to subtract the continuum could result in erroneous cross-correlations since spectra with similar continua -- but different lines -- can give stronger correlations than those with different continua but similar lines. 

Currently, continua have to either be modelled from population synthesis models \citep{Bruzual:2003,Panuzzo:2007}, requiring \emph{a priori} knowledge of galactic properties and physics, or they are computed from feature-deficient galactic spectra of a similar continuum type, which again requires the \emph{a priori} knowledge of how to identify and group galaxies which are of a similar type \citep{Koski:1976,Costero:1977}. Polynomial-fitting/statistical averaging methods are also frequently used when the noise is small enough so as not to conceal the continuum, or where denoising has already been employed, as in \citet{Stoughton:2002,Subbarao:2002}. In the very low SNR limit, it becomes exceedingly difficult to pinpoint exactly where the continuum lies, and polynomial fitting is not ideal.

In this paper, we introduce a new wavelet-based method that can isolate the continuum of a spectrum without having to defer to any knowledge of galaxy properties or physics, and that can operate at low SNR. We demonstrate on simulated data that this method performs well in both the low and high SNR regimes. Using a standard cross-correlation method for estimating the redshifts of galaxies, we demonstrate that an additional wavelet filtering to extract important features in each spectrum allows us to derive a selection criterion for galaxies for which we are able to accurately determine a redshift. This allows us to effectively clean our galaxy catalogue by removing catastrophic failures, resulting in a galaxy redshift catalogue with a guaranteed high success rate, and allowing us to accurately and confidently determine the redshifts of galaxies even in the very low SNR regime. This will be useful in large photometric surveys, such as the upcoming Euclid survey \citep{euclid,euclid2}, as the calibration of photometric redshifts requires spectroscopic redshift information with a very low incidence of catastrophic failures.

This paper is organised as follows. In section \sect{sec:pca}, we describe in detail the cross-correlation method used for redshift estimation. In \sect{sec:mock}, we briefly describe the construction of our datasets. In \sect{sec:wavelet}, we describe our wavelet-based continuum subtraction method, and the selection criteria for cleaning the catalogue. In \sect{sec:results}, we show results for various SNRs, using wavelength-independent white-Gaussian noise, and a mixed SNR catalogue with non-stationary (pixel dependent) Gaussian noise, and compare the proportion of catastrophic failures in cleaned catalogues to that obtained using the full spectral catalogue. In \sect{sec:real_noise}, we demonstrate the potential applicability of this software to real data. We demonstrate successful feature extraction on several real spectra obtained from the SDSS data archive. Lastly, in \sect{sec:conclusions} we summarise the key features of the Darth Fader algorithm, and identify potential future applications of the algorithm and the methods involved.

\section{Redshift Estimation by Cross-Correlation}\label{sec:pca}

To estimate galaxy redshifts, we employ a cross-correlation method similar to that described by \citet{Glazebrook:1998}. This method involves a cross-correlation of test galaxy spectra at unknown redshift with template spectra. 

We assume that any test spectrum $S^{'}_{\lambda}$ may be represented as a linear combination of template spectra $T_{i\lambda}$ ,

\begin{equation}\label{decomp}
S\,_{\lambda}^{'} = \sum_{i}\, a_{i}\,T_{i\lambda} \: ,
\end{equation}
where each template spectrum is normalised according to
\begin{equation}\label{normalisation}
\sum_{\lambda}\,T_{\lambda}^{2}=1 \: .
\end{equation}

If we choose to bin our spectra on a logarithmic wavelength axis, redshifting becomes proportional to a translation,
\begin{align}\label{lambdadef}
\Delta &= \log{(1+z)}\notag\\[6pt]
&= \log{\big( \lambda_{\, \textit{observed}} \, \big) } - \log{\big( \lambda_{\, \textit{rest frame}} \, \big) } \: .
\end{align}

The estimate of the goodness-of-fit between the template, now allowed to shift along the wavelength axis, and the test spectrum, at an unknown redshift, can be found by computing the minimum distance via a standard $\chi^2$, where the previous coefficients $a_{i}$ are now dependent upon redshift through $\Delta$,

\begin{equation}\label{chitemp}
\chi^{2}(\Delta) = \sum_{\lambda} \frac{w_{\lambda}^2}{\sigma_{\lambda}^{2}} \Big[ S_{\lambda} - \sum_{i} a_{i}(\Delta) \, T_{i(\lambda + \Delta)} \Big] ^{2} \: _{.}
\end{equation}

We can obtain the values of the expansion coefficients, $a_{i}$, by maximising equation \eqref{chitemp} with respect to $a_{i}$. Following the prescription in \citet{Glazebrook:1998}, we take the weighting function, $w_{\lambda}$, and the normally distributed errors, $\sigma_{\lambda}$, to be wavelength independent and constant, which gives

\begin{equation}\label{maximum}
a_{i}(\Delta) = \frac{\sum_{\lambda} S_{\lambda} \, T_{i(\lambda + \Delta)}}{\sum_{\lambda} T^{2}_{i(\lambda + \Delta)}} \: .
\end{equation}

The numerator in equation \eqref{maximum} is simply the cross-correlation of the galaxy spectrum with the $i^{th}$ template spectrum. Substituting back into equation \eqref{chitemp}, we obtain

\begin{equation}\label{chisub}
\chi^{2}(\Delta) \propto \sum_{\lambda} \Big[ \, S\,_{\lambda}^2 - \sum_{i} \, a_{i}^{2}(\Delta) \: T^{2}_{i(\lambda + \Delta)} \Big] \: _{.}
\end{equation}

For a large test catalogue that includes a variety of galaxy types, a large number of templates is needed to ensure the best match-up between template and test spectra. To use all of them in the cross-correlation would be excessively time-consuming. If it were possible to reduce the number of templates whilst still retaining most of the information content of these templates then we can render the method more practical.

Principal Component Analysis (PCA) is a simple tool that allows us to do just that: to reduce the dimensionality of this problem by extracting the most important features from our set of template spectra, the principal components. The general procedure involves the construction and subsequent diagonalisation of a correlation matrix to find eigenvectors and eigenvalues. It is possible to construct a correlation matrix either between the templates, or between the wavelength bins; the result is equivalent. We have chosen to do the correlation between the templates since in our case the number of templates is less than the number of wavelength bins, resulting in a smaller matrix that is simpler to manipulate:

\begin{equation}\label{corr}
C_{ij} = \sum_{\lambda} T_{i\lambda}\:T^{T}_{j\lambda} \: .
\end{equation}

Since this correlation matrix is always real and square-symmetric, it follows that it can be diagonalised,

\begin{equation}\label{diagon}
\mathbf{C} = \mathbf{R \Lambda R^{T}} \: ,
\end{equation}
where $\mathbf{\Lambda}$ represents the matrix of ordered eigenvalues (largest to smallest) and \textbf{R}, the matrix of correspondingly ordered eigenvectors. The eigentemplates, \textbf{E}, can then be obtained:

\begin{equation}\label{eigen}
E\,_{j\lambda} = \frac{ \sum_{i}\,R\,_{ij}^{T}\:T_{i\lambda}}{\sqrt{\Lambda_{j}}} \: ,
\end{equation}
with the resulting eigentemplates having the same dimensions as the original dataset, and satisfying the orthonormality condition

\begin{equation}\label{orthnorm}
\sum_{\lambda}\,E\,_{i\lambda}\:E\,^{T}_{j\lambda} = \delta\,_{ij} \: .
\end{equation}

The effect of PCA is that it re-orientates the dataset to lie along the orthogonal eigenvectors (axes) sorted by descending variance. It effectively creates an `importance order', such that the eigenvector with the greatest variance (largest eigenvalue) will tend to correspond to the strongest signal features of the untransformed dataset, with subsequent eigenvectors representing less significant signal features, and the final eigenvectors, with the smallest variances, representing noise. For example, if H$_{\alpha}$ is a very prominent feature in most of the template spectra, it will be present in one or more of the first few eigentemplates.

With this in mind we can now re-cast equation \eqref{decomp} in terms of an approximation of the sum of the first $N$ eigentemplates that are now allowed to be shifted along the wavelength axis:

\begin{equation}\label{eigendecomp}
S\,_{\lambda} \simeq \sum_{i=1}^{N} \, b_{i}(\Delta)\,E_{i(\lambda+\Delta)} \: ,
\end{equation}
where $b_{i}(\Delta)$ are new expansion coefficients for the new basis.

Using the orthogonality condition from equation \eqref{orthnorm}, equations \eqref{maximum} and \eqref{chisub} then become

\begin{equation}\label{bccf}
b(\Delta) = \sum_{\lambda} S_{\lambda} \, E_{i(\lambda + \Delta)} \: ,
\end{equation}
\\[-15pt]
\begin{equation}\label{chieigen}
\chi^{2}(\Delta) \propto \sum_{\lambda} \, S\,_{\lambda}^2 - \sum_{i=1}^{N} \, b_{i}^{2}(\Delta) \: .
\end{equation}

We then observe that the first term in equation \eqref{chieigen} is a constant in the $\chi^{2}$ function, and can be disregarded; therefore minimising the $\chi^2$ function in equation \eqref{chieigen} is equivalent to \emph{maximising} the related function, $\widetilde{\chi}\,^{2}$, defined as

\begin{equation}\label{lastchi}
\chi^{2}(\Delta) \sim \widetilde{\chi}\,^{2}(\Delta) = \sum_{i=1}^{N} \, b_{i}^{2}(\Delta) \: .
\end{equation}

Hence, $\widetilde{\chi}\,^{2}(\Delta)$ is computed by first computing the cross-correlation of each of the $N$ retained eigentemplates $E_i$ with the galaxy spectrum (equation \eqref{bccf}), and then summing $b_{i}^{2}(\Delta)$ over these eigentemplates. We can further simplify the problem by noting that a convolution between two real signals transforms into a multiplication in Fourier space between the individual Fourier transforms of the galaxy and non-redshifted template spectra, with the advantage that $\Delta$ becomes a free parameter. Hence we obtain

\begin{equation}\label{fourierb}
b_{i}(\Delta) =\mathcal{F}^{-1} \big( \hat{S}_{k} \, \hat{E}_{ik} \big)= \frac{1}{M} \sum_{k=0}^{M-1} \hat{S}_{k} \, \hat{E}_{ik} \, e^{\frac{2 \pi \mathbbm{i} k \Delta}{M}} \: ,
\end{equation}
and
\begin{equation}\label{chifourier}
 \widetilde{\chi}\,^{2} (\Delta) = \sum_{i=1}^{N} \Big[ \mathcal{F}^{-1} \big( \hat{S}_{k} \, \hat{E}_{ik} \big) \Big]^{2} \: ,
\end{equation}
where $\hat{S}_{k}, \, \hat{E}_{ik}$, represent the Discrete Fourier Transforms (DFTs) of $S_{\lambda}, \, E_{i\lambda}$; and $\mathbbm{i}, \, \mathcal{F}^{-1}$ represent $\sqrt{-1}$ and the inverse DFT respectively.

Now that we have obtained equation \eqref{chifourier} it is an easy task to extract the estimate for the redshift, $z$. The $\widetilde{\chi}\,^{2}$ function reaches a maximum when the shift of the templates along the log-wavelength axis corresponds to the true shift of the galaxy spectrum, so that the redshift is estimated to be where $\Delta = \Delta_{\widetilde{\chi}} \, (= \Delta|_{\widetilde{\chi}=\widetilde{\chi}_{max}})$, giving

\begin{equation}\label{zest}
z_{est} = 10^{\, \delta_{s} \, \Delta_{\widetilde{\chi}}}-1 \: ,
\end{equation}
where $\delta_{s}$ is the grid spacing on the $\log_{10}$-wavelength axis.

Note that, for this PCA/cross-correlation redshift estimation method, both the template and galaxy spectra must be free of continuum. This is important to ensure that it is the spectral features from each spectrum that are being matched to one another, rather than to any continuum features, which may lead to spurious correlations and hence confusion in the determination of the galaxy redshift. In Darth Fader, we use an entirely empirical method for subtracting the continuum that is based on the wavelet decomposition of the spectrum, and which is easily automated and very effective regardless of the signal-to-noise of the spectrum under consideration. This method will be described in detail in \sect{sec:modelling}.

\section{Simulations}\label{sec:mock}

The redshift estimation method described above requires two separate spectral catalogues: a set of galaxy spectra with noise and covering a range of redshifts that we aim to estimate (the test catalogue), and a set of noise-free, zero-redshift template spectra. We use the CMC set of simulations as provided by \citet{Jouvel:2009,Zoubian:2013}, which are based on the observed COSMOS SEDs of \citet{Ilbert:2009,Capak:2009}. We then rebin a randomly selected sub-sample of the CMC master catalogue onto an evenly spaced  $\log_{10 \;}\lambda$ wavelength grid, spanning the range between \numprint{3000}\,\textup{\AA} to \numprint{10500}\,\textup{\AA} for the test catalogue, with a wider range for the template spectra of \numprint{3000}\,\textup{\AA} to \numprint{20900}\,\textup{\AA}.

The template catalogue was compiled by randomly selecting galaxies within the simulation set with redshift less than $z=0.1$. The number of galaxies selected was chosen to be roughly 10\% of the size of the main test catalogue, in order to ensure a representative sample of template galaxies. This resulted in a template catalogue of 277 simulated spectra, which were then blueshifted to be at zero redshift, and a test catalogue of \numprint{2860} simulated spectra with redshifts in the range $0.005<z<1.7$.

This choice of binning, and a similar pixelisation scheme as in \citet{Smee:2012}, gives a constant resolution across the spectrum of $R \, \big( = \sfrac{\lambda}{\Delta\lambda}\big) \sim 850$ for all the catalogues, and a grid spacing of $\delta_{s} = 2.17 \times10^{-4} \, \log_{10} \textup{\AA}$; as compared to SDSS where the resolution and grid spacing are $R\sim$ \numprint{1845}, and $\delta_{s} = 1.0 \times10^{-4} \, \log_{10} \textup{\AA}$ respectively.

The template spectra are required to have a larger wavelength span than the test spectra since they must be able to accommodate a large enough overlap to identify the correct (global) cross-correlation minima with these test spectra at all redshifts under consideration. A restricted wavelength span on the set of template spectra will necessarily reduce the maximum redshift at which you can cross-correlate. This frequently will result in the cross-correlation picking a local minimum that exists in the overlap -- since the global minimum lies outside this overlap -- often resulting in a confusion between one principal feature for another.

Wavelength-independent (white) Gaussian noise was then added to the test catalogue to generate several catalogues in which all the galaxies had the same SNR. We define our SNR in the same manner as in the SDSS pipeline \citep[][for BOSS/SDSS III]{Bolton:2012}\footnote{The idlspec2d pipeline software package is available at: \url{http://www.sdss3.org/dr8/software/products.php}}, relative to the median SNR in the SDSS r-band filter \citep{Fukugita:1996}:

\begin{equation}\label{snrdef}
\mathrm{SNR}\,_{r} = median\Big[\,\frac{flux}{\sigma}\,\Big]\,^{\numprint{6760}\,\textup{\AA}}_{\numprint{5600}\,\textup{\AA}} \: ,
\end{equation}
where $\sigma$ is the standard deviation of our added white-Gaussian noise, and the subscript r denotes that the median is calculated between the bounds of SDSS r-band filter (\numprint{5600}\,\textup{\AA} to \numprint{6760}\,\textup{\AA}).

We choose this particular definition of SNR so as to be consistent with a realistic survey such as SDSS. The specific choice of SDSS band on which to base the SNR definition does not affect the method presented in this paper, however the motivation for choosing the r-band over any of the other SDSS bands is fully explained in \citet{Strauss:2002}. Whilst this definition of SNR is a good proxy for SNR on the continuum, and as such allows a simple comparison between different spectra, it should be cautioned that it is not necessarily a good proxy for the SNR on specific features.

An additional mixed SNR catalogue was generated by adding pixel-dependent Gaussian noise such that the spectra in the catalogue had a uniform distribution in SNR in the range $1.0< \mathrm{SNR} < 20$ as described in section \sect{sec:real_noise}.

\section{The Darth Fader Algorithm}\label{sec:wavelet}

Darth Fader works in a number of different steps to take a test catalogue of galaxy spectra containing both spectral lines and continuum features (as well as noise), and to output a clean catalogue of galaxies for which we are able to obtain robust and accurate redshift estimates via PCA and cross-correlation. A schematic of the full algorithm is shown in Figure \ref{fig:algorithm}.

\begin{figure*}
 \centering
 \vspace{60pt}

\tikzstyle{decision} = [diamond, draw, fill=red!20,
    text width=4.5em, text badly centered, node distance=2.5cm, inner sep=0pt]
\tikzstyle{block} = [rectangle, draw, fill=blue!20,
    text width=5em, text centered, rounded corners, minimum height=4em]
\tikzstyle{line} = [draw, very thick, color=black!50, -latex']
\tikzstyle{cloud} = [draw, ellipse,fill=green!20, node distance=2.5cm, text centered, 
    minimum height=2em]

\begin{tikzpicture}[scale=0.5, node distance = 2cm, auto]

    % Place nodes
    \node at (167pt,100pt) [align=center](title){\textbf{\large The Darth Fader Algorithm}};
    \node [cloud,align=center] (temp) {template\\catalogue};
    \node [cloud, right of=temp, node distance=6cm,align=center] (test) {test\\catalogue};
    \node [block, below of=temp, node distance=4cm ] (contsub) {blind continuum subtraction};
    \node [block, below of=test] (noise) {addition of noise};
    \node [block, below of=noise] (contsub2) {blind continuum subtraction};
     \node [block, below of=contsub] (pca) {PCA decomposition};
     \node [cloud, below of=pca, node distance=2cm,align=center] (eigen) {retain N\\eigentemplates};
      \node [block, below of=contsub2] (fdr) {FDR denoising};
       \node [block, below of=eigen] (ccf) {cross-correlation};
    \node [decision, below of=fdr,node distance=4cm] (peaks) {Does the denoised spectrum possess 6 total features or more?};
    \node [block, below of=ccf, node distance=4cm] (z) {estimate redshift};
     \node [block, below of=peaks, node distance=4cm] (discard) {discard spectrum};
    %% Draw edges
    \path [line] (temp) -- (contsub);
    \path [line] (test) -- (noise);
    \path [line] (noise) -- (contsub2);
    \path [line] (contsub) -- (pca);
    \path [line] (peaks) -- node [near start, color=black] {yes} (ccf);
     \path [line] (peaks) -- node [near start, color=black] {no} (discard);
    \path [line,dashed] (contsub) -- node [color=black] {identical procedure} (contsub2);
   \path [line,dashed] (contsub2) -- (contsub);
   \path [line] (pca) -- (eigen);
   \path [line] (contsub2) -- (fdr);
   \path [line] (fdr) -- (peaks);
   \path [line] (eigen) -- (ccf);
   \path [line] (ccf) -- (z);

\end{tikzpicture}\vspace{60pt}
\capt{This figure illustrates the operation of Darth Fader. The number of eigentemplates to be retained is at the discretion of the user, and may depend on the distribution of spectral types in the data. We have chosen to retain 20 eigentemplates because they encompass 99\% of the total eigenvalue weight. In general the number retained will be significantly less than the number of spectra in the original template catalogue. The FDR denoising procedure denoises the positive and negative halves of the spectrum independently, with positivity and negativity constraints respectively. The requirement of six features or more in the denoised spectrum effectively cleans the catalogue of galaxies likely to yield catastrophic failures in their redshift estimates. It should be noted that a `no' decision represents the termination of that spectrum from the test catalogue and our analysis. Potentially, a further analysis could be employed at this point in order to review spectra possessing 5 features, for example, to see if further information, such as the association of standard lines to these features, can yield a redshift estimate (which could be used in conjunction with an estimate obtained from cross-correlation, with corresponding redshifts likely to be correct).}\vspace{80pt}
\label{fig:algorithm}
\end{figure*}

In order to estimate the redshift of galaxies by cross-correlation, both the templates and the test galaxy spectra must be continuum-free. Current methods for continuum subtraction rely on a handful of principal techniques: careful modelling of the physics of galaxies to estimate the continuum, a matching of continua between featureless galaxy spectra (typically sourced from elliptical galaxies or galactic bulges) and the spectra from which we wish to remove the continuum, or a polynomial fitting. \citep{Koski:1976,Costero:1977,Panuzzo:2007}. Median filtering to remove spectral features is also a possibility; however this relies on a fixed choice for the median filtering window size, which may not be appropriate for resolved lines or blended line doublets.

The first two of these methods have the disadvantage of requiring some knowledge of galaxy physics (which may not be precisely known), and being somewhat restricted to lower redshift/higher SNR galaxies. Careful modelling is computationally intensive and liable to result in failure if unusual galaxy types are found, or if the physics involved is not fully understood or modelled well enough. Continuum-matching methods require \emph{a priori} knowledge of the galaxy type of a set of galaxies, and/or are reliant on the correspondence between similar looking continua (with one relatively featureless, in order to remove one from the other) which may not exist for all spectra. All matching methods have a problem at high noise levels where different but superficially similar spectra are mistakenly associated. Polynomial fitting is a a further alternative that is limited to high signal-to-noise spectra, or spectra that have been denoised beforehand (as applied to the SDSS data release, \citet{Stoughton:2002}). 

By contrast our new method of continuum subtraction is completely empirical, requires no knowledge of the physics or type of the galaxy involved and can be used even with very noisy spectra. This method relies on a \textbf{multiscale} modelling of the spectra, as described below.

\subsection{Spectra modelling}
\label{sec:modelling}

We model the galaxy spectrum as a sum of three components -- continuum, noise and spectral lines:

\begin{equation}
\label{LNC}
S = L + N + C \, ,
\end{equation}
where $L$ contains the spectral line information, $N$ is the noise and $C$ is the continuum.  $L$ can be decomposed into two parts, emission lines $L_e$ and absorption lines $L_a$: 
$ L = L_e + L_a$, where, provided the continuum has been removed, $L_e > 0$ and $L_a < 0$.

The problem is then to estimate these components, $L$ and $C$, from a unique data set. This is possible assuming that features of $L$ are important only on small and intermediate scales, while the continuum contains no small scale features, and is dominant on large scales only. However, several problems remain:

\begin{itemize}
\item Strong emission/absorption lines impact significantly at all frequencies, so a low pass filtering of the input is not sufficient to properly estimate the continuum, $C$.\vspace{4pt}
\item Both emission and absorption lines are difficult to detect because of the presence of noise.
\end{itemize}

\subsection{Continuum removal}

The method we present in the following is done in four steps, two for continuum estimation and two for absorption and emission line estimation.

\begin{enumerate}
\item We first detect strong emission and absorption lines, which could be seen as outlier values for continuum emission.
\item We subtract from the data these specific strong features and estimate the continuum from this.
\item We then re-estimate the emission lines from the original data now continuum-subtracted via steps 1 and 2.
\item We also re-estimate the absorption lines in a similar way.
\end{enumerate}

\subsubsection{Strong line removal using the pyramidal median transform}

In order to detect strong emission and absorption lines that could be seen as outliers for the continuum, we need a tool that is highly robust to these outliers. The choice of median filtering is generally the correct one for such a task. However, fixing the median filtering window size to the width of an unresolved line is not appropriate for blended line doublets and resolved lines. In our case, a better choice therefore is the multiscale median transform that was proposed for cosmic ray removal in infrared data \citep{Starck:1996_tech,Starck:2006}. Furthermore its pyramidal nature allows us to significantly speed up computation time \citep{Starck:1996}. In this framework, strong features of different width can be efficiently analysed.

In a general multiscale transform\footnote{IDL routines to compute this and other wavelet transforms are included in the \textbf{iSAP} package available at: \url{http://www.cosmostat.org/software.html}}, a spectrum of $n$ bins, $S_{\lambda} = S \, [1,\dots,n]$ can be decomposed into a coefficient set, $W = \{w_1, \dots, w_J, c_J\}$,  as a superposition of the form

\begin{equation}\label{starlet}
S_{\lambda} = c_{J}(\lambda) + \sum_{j=1}^{J} w_j(\lambda) \;,
\end{equation}
where $c_{J}$ is a smoothed version of the original spectrum $S_{\lambda}$, and the $w_j$ coefficients represent the details of $S_{\lambda}$ at scale $2^{-j}$; thus, the algorithm outputs $J+1$ sub-band arrays each of size $n$. The present indexing is such that $j = 1$ corresponds to the finest scale or highest frequencies.

We use a similar multiscale transform in the following algorithm for strong line detection:

\begin{itemize}
\item take the pyramidal median transform (PMT) of the input spectrum $S$ (a median window of size 5 was used in all our experiments), we get a set of bands $w_{j}$ and $c_J$ at different scales $j$,  $\mathcal{P}(S) = \{w_1, \dots, w_J, c_J\}$. Where $w_{j}$ corresponds to multiscale median coefficients, and can be interpreted as the information lost 
between two resolutions when the downgrading is performed using the median filtering followed by a downsampling of factor 2. The $c_J$ term corresponds to a very coarse resolution of the input signal. Full details can be found in \citet{Starck:2010}.
\item for each band $w_j$, threshold all coefficients with an amplitude smaller than four times the noise level.
\item set the coarse resolution, $c_J$, to zero.
\item reconstruct the denoised spectrum $S_1$.
\end{itemize}

$S_1$ represents a crude estimation of the lines $L$, mainly because the noise behaviour in the pyramidal decomposition cannot be calculated as well as in a linear transform such as the Fourier or the wavelet transform. However the process is much more robust than with a linear transform since strong lines with small width will not contaminate the largest scales as it would be the case for instance with wavelets, resulting in artefacts termed `ringing'.

Since $S_1$ contains the signal from strong lines, $S_2 (= S - S_1)$ will be free of any strong features, and robust continuum estimation can easily be derived from it. 

\subsubsection{Continuum extraction}
\label{sec:dwt}

The second step is therefore to estimate the continuum from $S_2$. The largest scale of  $S_2$ should contain the continuum information (see first term in equation \eqref{starlet}), whilst the noise and undetected lines are expected to be dominant on smaller scales. So now the coarsest scale in a wavelet decomposition, or any low pass filtering, would give us a good estimation for the continuum. The great advantage of wavelets for this task, as compared to a simple low pass filtering performed in Fourier space for example, is to allow a greater flexibility for handling the border (i.e. minimising edge effects), and there being no requirement to assume periodicity of the signal. We do this using the starlet wavelet transform, also called isotropic undecimated wavelet transform (equation \eqref{starlet}).

This transformation is simply a new representation of the original signal, which can be recovered through a simple summation. For a detailed description of the starlet transform see \citet{Starck:2010}, which has further been shown to be well-adapted to astronomical data where, to a good approximation, objects are  commonly isotropic \citep{Starck:1994,Starck:2006}. 

We therefore estimate the continuum by first taking the wavelet transform of $S_2$, i.e.: $W^{[S_2]}   = \{w^{[S_2]}_1, \dots, w^{[S_2]}_J, c^{[S_2]}_J\}$, and then retaining only the largest scale: $c^{[S_2]}_J = C$. This continuum can now be subtracted from the original noisy spectra to yield a noisy, but now continuum-free, spectrum.

\subsubsection{Example}

\begin{figure*}[h!]

 \centering
  \includegraphics[trim=2cm 0cm 1.5cm 1.5cm, clip=true,width=0.9\textwidth]{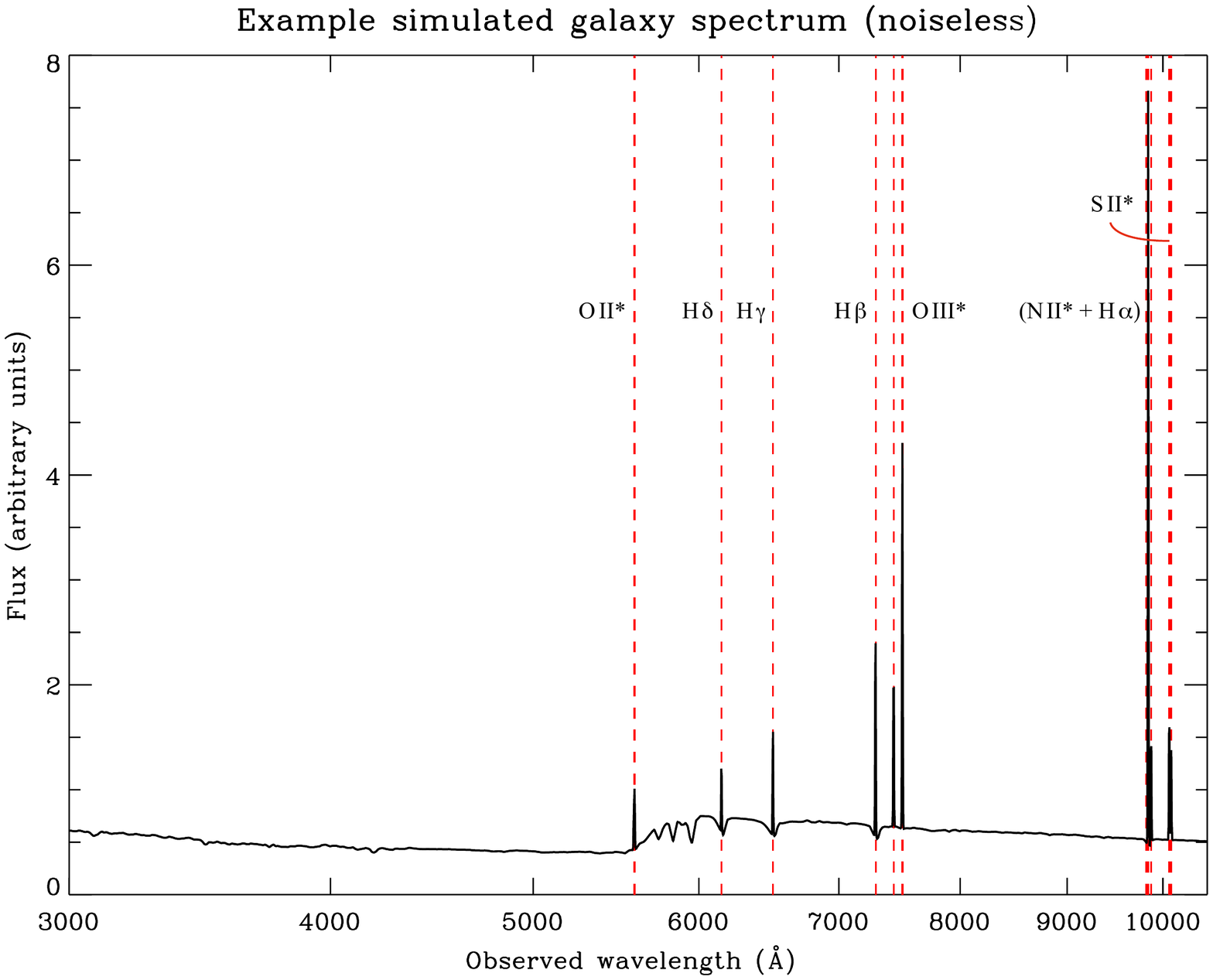}\vspace{-25pt}

 \capt{This figure shows an example spectrum from the test catalogue ($z=1.4992$), prior to the addition of noise. The main emission lines are labeled; with an asterisk denoting a doublet feature. The [O$\,$II] doublet is fully blended in this spectrum.}
\label{fig:purespec}
\end{figure*}

\begin{figure*}[h!]
\centering
       
         \begin{subfigure}[b]{0.49\textwidth}
                \includegraphics[trim=2cm 0cm 1.5cm 1.5cm, clip=true,width=\textwidth]{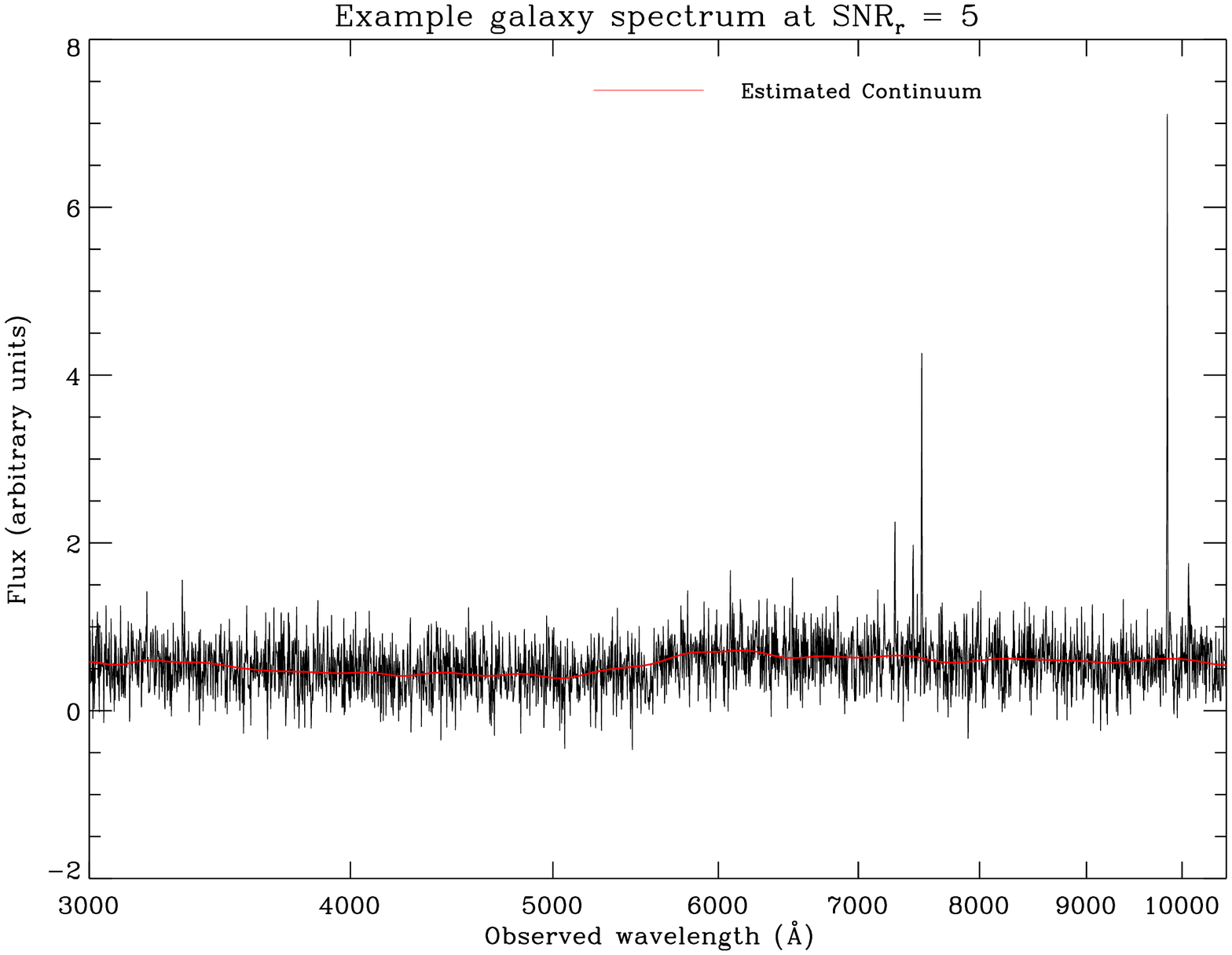}\vspace{-20pt}
                \caption{}
                \label{fig:noiselow}
              \end{subfigure}
              ~
              \begin{subfigure}[b]{0.49\textwidth}
                \includegraphics[trim=2cm 0cm 1.5cm 1.5cm, clip=true,width=\textwidth]{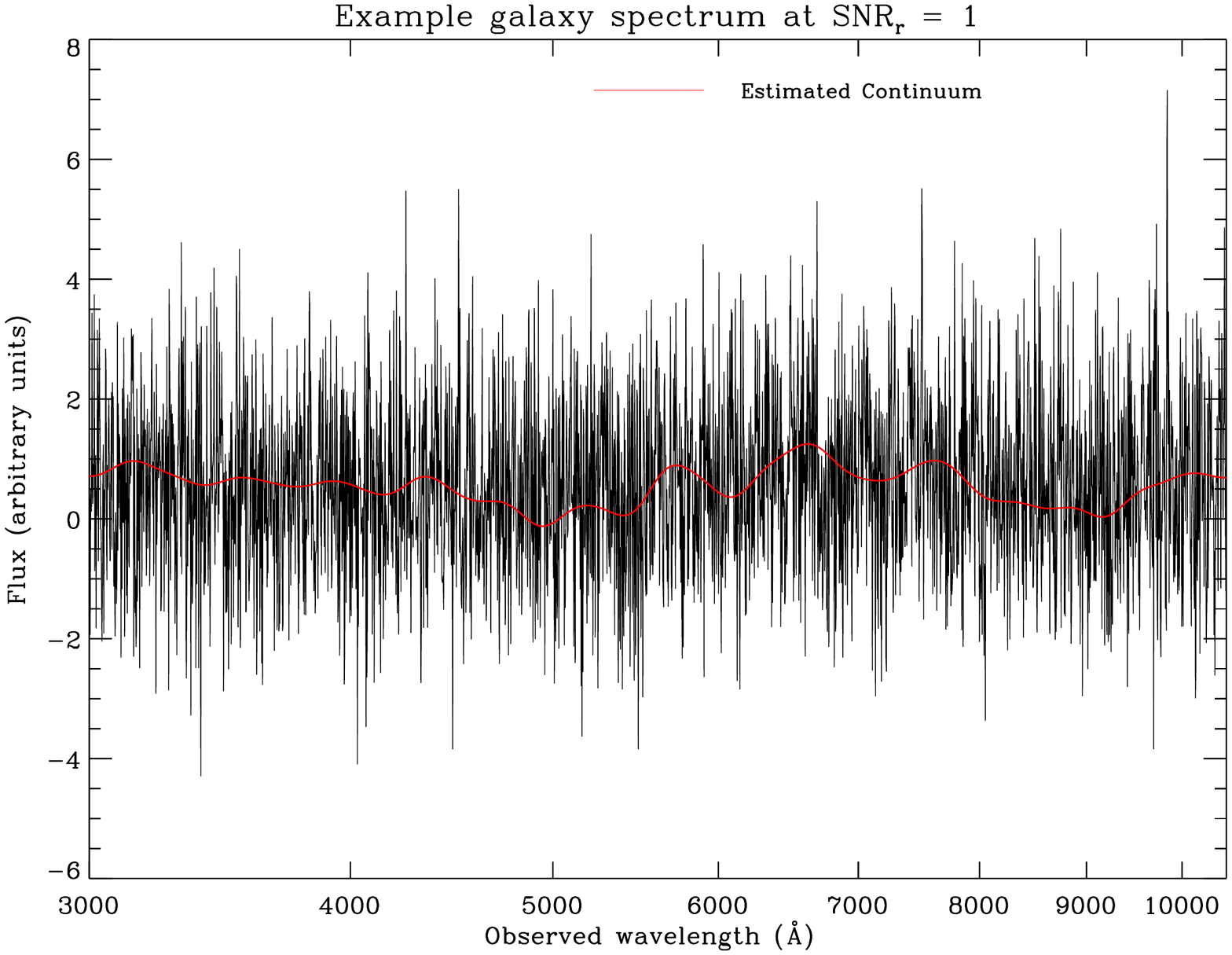}\vspace{-20pt}
                \caption{}
                \label{fig:noisehigh}
              \end{subfigure}
              \caption{This figure shows a same spectrum as that in figure \ref{fig:purespec} but with manually added white-Gaussian noise at a signal-to-noise level in the r-band of 5 in figure \ref{fig:noiselow}, and of 1 in figure \ref{fig:noisehigh}. The red lines indicate the empirically-determined continua in each case. Many of the prominent lines are easily visible by eye at the higher SNR of 5, whereas at the lower SNR of 1 most of the lines are obscured, with only H$_{\alpha}$ being sufficiently prominent so as to be detectable. The continuum estimate is good at the SNR of 5, and comparatively poor, but of the correct order of magnitude, at the lower SNR due to the dominating influence of noise. As an indication of \emph{line-SNR}, we quote the values for the SNR on H$_{\alpha}$ for these particular spectra as 8.9 and 1.7 respectively for figures \ref{fig:noiselow} and \ref{fig:noisehigh}.}
              \label{fig:modelspec}
\end{figure*}

We show in figure \ref{fig:purespec} an example noise-free spectrum from our simulated catalogue, containing both line features and continuum. In figures \ref{fig:noiselow} and \ref{fig:noisehigh}, we show the same spectrum with noise added resulting in values of 5 and 1 for the  SDSS r-band SNR. We note that galaxy surveys select galaxies based on their SNR, typically with a lower bound at an SNR of 5-10 on the continuum. Over-plotted in these latter two figures is the continuum as estimated by the method described above, for the spectrum with an SNR of 5, the continuum fit can be seen to be quite good. {At lower SNR, the continuum fit is quite poor, as $S_1$ is poorly estimated for this particular noise realisation, and the dominating influence of noise effectively conceals the continuum.} However, this continuum estimate at low SNR is still well within the noise, and the correct order of magnitude. For reference we calculate the SNR on the H$_{\alpha}$ line in each case by taking the ratio of the mean flux per pixel on the line and the noise standard deviation; the H$_\alpha$ SNR was found to be 8.9, and 1.7 respectively for r-band SNRs 5 and 1.

\subsection{Absorption/emission line estimation using sparsity}
\label{sec:wavcoef}

The wavelet representation of a signal is useful because it enables one to extract features at a range of different scales. In many cases, a wavelet basis is seen to yield a sparse representation of an astrophysical signal (such as a spectrum or a galaxy image), and this sparsity property can be used for many signal processing applications, such as denoising, deconvolution, and inpainting to recover missing data \citep[e.g.][]{Fadili:2009,Starck:2010}. In this paper, we focus on one such application: that of denoising spectra using wavelet filtering.

The basic idea underlying sparse wavelet denoising is that the signal we are aiming to recover is sparsely represented in our chosen wavelet dictionary. This means that the signal is completely represented by a small number of coefficients in wavelet space.\footnote{This is analogous to the representation of periodic signals in Fourier space, where they may be represented by only a few frequencies in this domain.} This sparsity property means that if we are able to identify the important coefficients, it is straightforward to extract the signal from the noise. 

There are various methods to do this; one simple method would be $K\sigma$ clipping, where a threshold is set relative to an estimate of the noise, and all coefficients with an SNR less than $K$ are set to zero. A more sophisticated method involves the use of a False Discovery Rate (FDR) threshold, which allows us to control contamination from false positive lines arising from noise features. This method will be described in detail below. Wavelet denoising has been previously applied successfully to both stellar (\citealt{Fligge:1997,Lutz:2008}) and galactic spectra (\citealt{Stoughton:2002}, for the SDSS early data release).

\subsubsection{Sparse Wavelet Modelling of Spectra}
\label{sec:modelling}

As the continuum $C$ is now estimated, we will now use the continuum free spectrum $S_c = S - C $. We can consider now that the remaining problem is to estimate the lines, assuming $S_c =  L +  N $ and $L = L_e + L_a$.
We exploit the wavelet framework in order to decompose it into two  components:  line features, and noise. 
This is done using a modified version of a denoising algorithm based on the Hybrid Steepest Descent (HSD) minimisation algorithm developed by \citet{Yamada:2001}.

Hence, we can reconstruct $L$ by solving the following optimisation problem:

\begin{equation}
\label{L1min}
 \min_{L} \norm{\mathbf{\hat{\mathcal{W}}} L}_1, \quad \st \quad S \in \mathcal{C}, 
\end{equation}
where $\mathbf{\hat{\mathcal{W}}}$ is the wavelet transform operator, $\norm{.}_1$ is the $\ell_1$ norm, which promotes sparsity in the wavelet domain, and $\mathcal{C}$ is convex set of constraints, the most important of which is a linear data fidelity constraint:
\begin{equation}
\label{constraint}
 \abs{ { w_j^{[S]}(\lambda) - w_j^{[L]}(\lambda)}} \le \varepsilon_{j} , \; \forall \; (j,\lambda) \in \mathcal{M} \; . 
\end{equation}
Here $w_j^{[S_c]}$ and $w_j^{[L]}$ are respectively the wavelet coefficients of $S_c$ and $L$, and $\varepsilon_{j}$ is an arbitrarily small parameter that controls how closely the solution $L$ matches the input data. The constraint set $\mathcal{C}$ may also include further constraints, such as positivity for emission line-only spectra, etc. Note that the large scale coefficients $c_J$ are not considered in this minimisation, as we do not expect the largest scales to contain any useful information since the continuum has been subtracted. $\mathcal{M}$ is the \emph{multiresolution support} \citep{Starck:1995}, which is determined by the set of detected significant coefficients at each scale $j$, and wavelength $\lambda$, as

\vspace{-6pt}
\begin{equation}
\mathcal{M}:= \{(j,\lambda)\ |\ \textrm{if}\ w_{j}(\lambda)\ \textrm{is declared significant}\} \; . %\vspace{5pt}
\end{equation} 

The multiresolution support is obtained from the noisy data $S_{c}$ by computing the forward transform coefficients $W = \{w_1, \dots, w_J, c_J\}$, and recording the coordinates of the coefficients $w_{j}$ with an absolute value larger than a detection level threshold $\tau_{j}$, often chosen as $\tau_{j} = K\sigma_{j,\lambda}$, where $K$ is specified by the user (typically between 3 and 5) and $\sigma_{j,\lambda}$ is the noise standard deviation at scale $j$ and at wavelength $\lambda$. When the noise is white and Gaussian, we have $\sigma_{j,\lambda} = \sigma_{j}$, and $\sigma_{j}$ can directly be derived from the noise standard deviation in the input data. When the noise is Gaussian, but not stationary, which is generally the case for spectral data, we can often get the noise standard deviation per pixel $\sigma_{\lambda}$ from the calibration of the instrument used to make the observation, and $\sigma_{j,\lambda}$ can be easily derived from $\sigma_{\lambda}$ \citep{Starck:2006}.

An interesting and more efficient alternative to this standard $K\sigma$ detection approach is the procedure to control the False Detection Rate (FDR). The FDR method \citep{Benjamini:1995}\footnote{\citeauthor{Benjamini:1995} term FDR as false \emph{discovery} rate in their paper; it is exactly analogous to what we term false detection rate in this paper.} allows us to control the average fraction of false detections made over the total number of detections. It also offers an effective way to select an adaptive threshold, $\alpha$.

In the most general context, we wish to identify which pixels of our galaxy spectrum contain (predominantly) signal, and are therefore `active', and those which contain noise and are therefore `inactive'. The measured flux in each pixel, however, may be attributed to either signal or noise, with each having an associated probability distribution. When deciding between these two competing hypotheses, the null hypothesis is that the pixel contains no signal, and the alternative hypothesis is that signal is present (in addition to noise).

The FDR is given by the ratio:

\begin{equation}
FDR = \frac{V_{f}}{V_{a}}\ ,
\end{equation}
where $V_{f}$ is the number of pixels that are truly inactive (i.e. are part of the background/noise) but are declared to be active (falsely considered to be signal), and $V_{a}$ is the total number of pixels declared active.

The procedure controlling the FDR specifies a fractional threshold, $\alpha$, between 0 and 1 and ensures that, \emph{on average}, the FDR is no bigger than $\alpha$:

\begin{equation}
\langle FDR \rangle \leq \frac{V_{i}}{V_{T}}.\alpha \leq \alpha\ .
\end{equation}

The unknown factor $\sfrac{V_{i}}{V_{T}}$ is the proportion of truly inactive pixels; where $V_{i}$ is the number of inactive pixels, and $V_{T}$ the total number of pixels.

A complete description of the FDR method can be found in \citet{Starck:2006} and, from an astrophysical perspective, in \citet{Miller:2001}. FDR has been shown to outperform standard methods for source detection \citep{Hopkins:2002}, and \citet{Pires:2006} have shown that FDR is very efficient for detecting significant wavelet coefficients for denoising of weak lensing convergence maps. In this paper, the FDR method is applied at each wavelet scale, and hence gives a detection threshold $\tau_j$ per wavelet scale.
 
The minimisation in equation \eqref{L1min} can be achieved using a version of the Hybrid Steepest Descent (HSD) algorithm adapted to non-smooth functionals, full details of which can be found in \citet{Starck:2010}.

In practice, we separately estimate emission lines and absorption lines by running the algorithm twice, first with a positivity constraint to get $L_e$, and then with a negativity constraint to estimate $L_a$. We found this approach more efficient than a single denoising without constraint, allowing us to reduce ringing (a type of denoising artefact that would compound feature counting) around detected lines well. Our final estimate of $L$ is then obtained by $L = L_e + L_a$.

\subsection{Example}

\begin{figure*}[h!]

 \centering
  \includegraphics[trim=2cm 0cm 1.5cm 1.5cm, clip=true,width=0.9\textwidth]{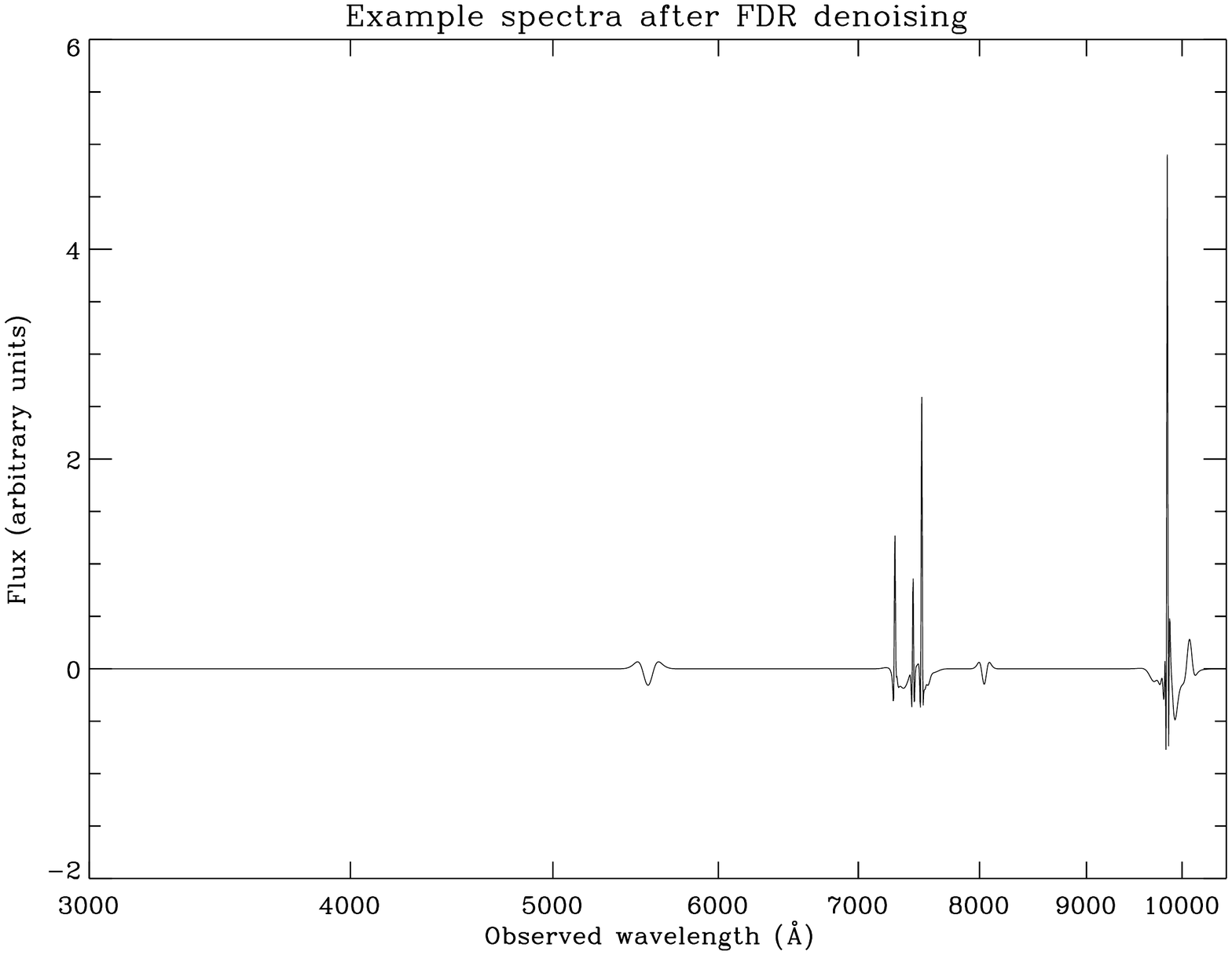}\vspace{-25pt}

 \capt{This figure is the result of an unrestricted denoising of the spectrum in figure \ref{fig:noiselow} with an FDR threshold corresponding to an allowed rate of false detections of $\alpha = 4.55\%$. The  [O$\,$III] doublet, H$_{\alpha}$ and H$_{\beta}$ are all cleanly identified. There are small features corresponding to [O$\,$II] and [S$\,$II], and a spurious feature at just over \numprint{8000}\,\textup{\AA}. The FDR denoising of \ref{fig:noisehigh} fails to detect any features for this particular spectrum, noise-realisation and choice of FDR threshold, and thus returns a null spectrum (not shown).}
\label{fig:denoise}
\end{figure*}

As an example, we show in figure \ref{fig:denoise} the first attempt at the reconstruction of the lines, $L$, from figure \ref{fig:noiselow}, using an FDR threshold of $\alpha = 4.55\%$. Here, the positive and negative halves of the spectrum have \emph{not} received independent treatment and the denoising is unrestricted since it is for the purpose of continuum-subtraction. It is the FDR denoising with the aim of feature-counting (figure \ref{fig:fclow}) that requires a separate treatment of positive and negative halves of the spectrum; the FDR denoising in order to isolate the continuum does not require this procedure. The denoising of figure \ref{fig:noisehigh} fails to detect any features, and thus returns a null spectrum.

\begin{figure*}[h!]
\centering
       
         \begin{subfigure}[b]{0.49\textwidth}
                \includegraphics[trim=2cm 0cm 1.5cm 1.5cm, clip=true,width=\textwidth]{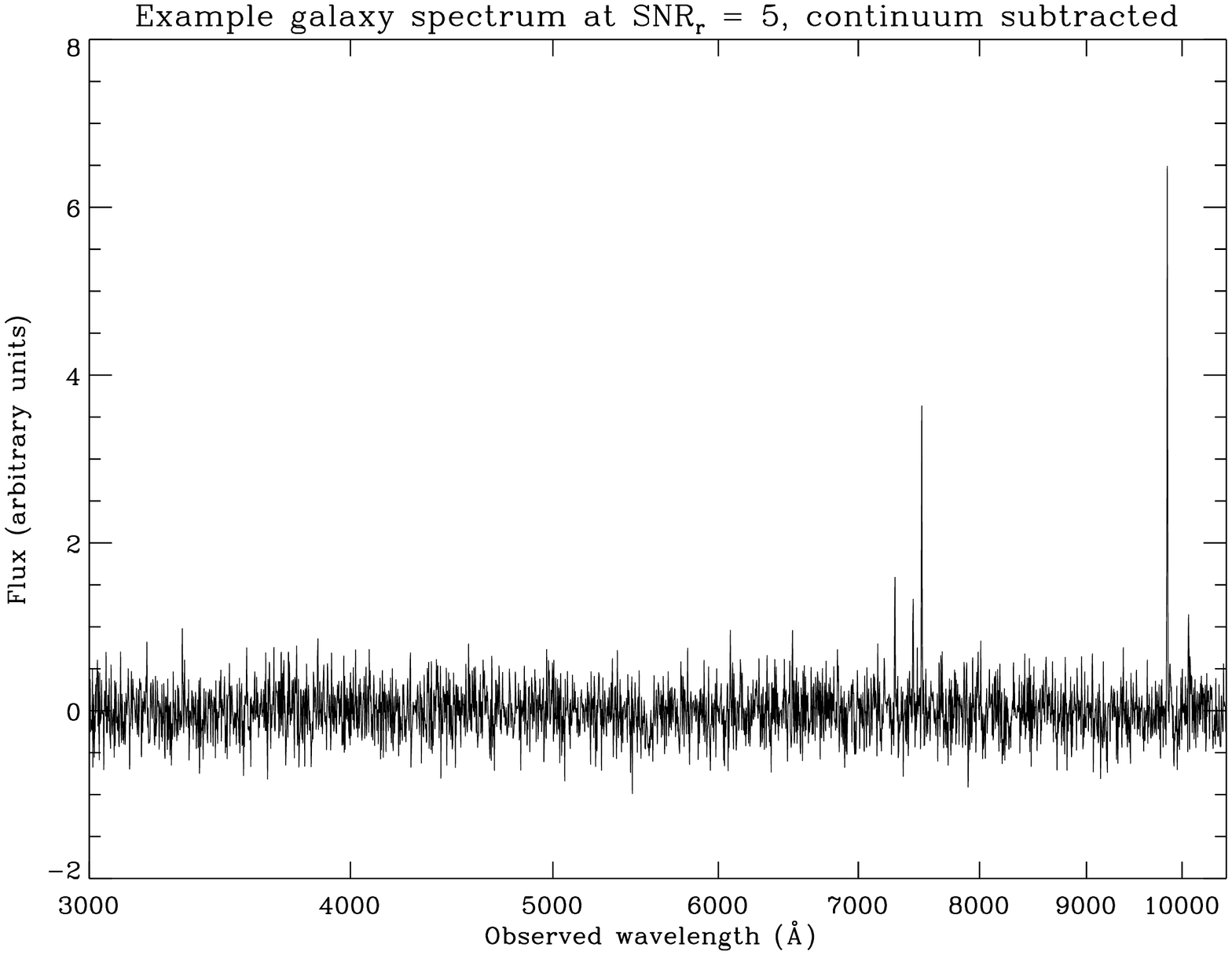}\vspace{-20pt}
                \caption{}
                \label{fig:contnoiselow}
              \end{subfigure}
              ~
              \begin{subfigure}[b]{0.49\textwidth}
                \includegraphics[trim=2cm 0cm 1.5cm 1.5cm, clip=true,width=\textwidth]{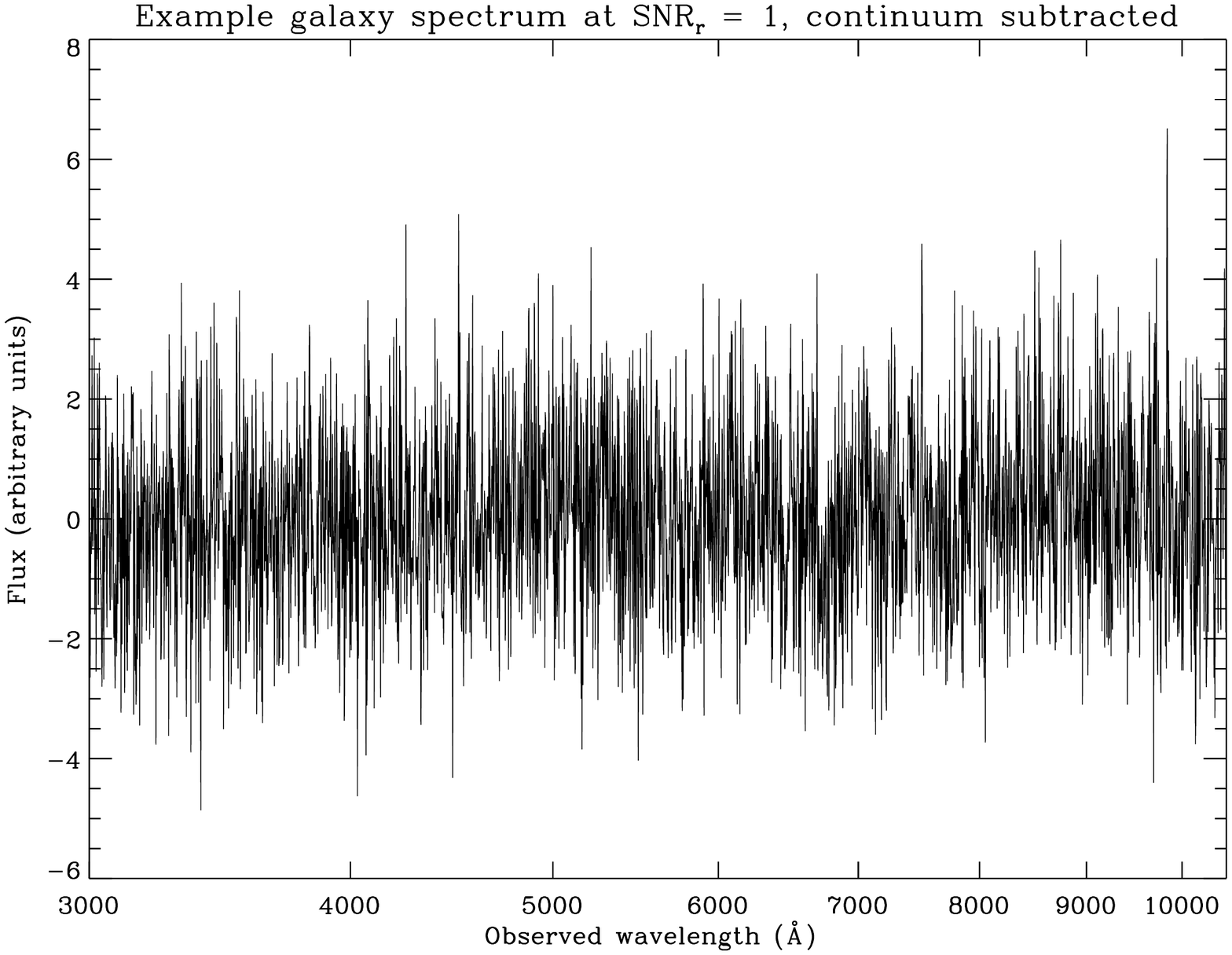}\vspace{-20pt}
                \caption{}
                \label{fig:contnoisehigh}
              \end{subfigure}
              \caption{Figures \ref{fig:contnoiselow} and \ref{fig:contnoisehigh} are the spectra as shown in figures \ref{fig:noiselow} and \ref{fig:noisehigh} with their empirically determined continua subtracted.}\label{fig:nocontspec}
\end{figure*}

\begin{comment}
\begin{figure}[h!]

 \centering
  \includegraphics[trim=2cm 0cm 1.5cm 1.5cm, clip=true,width=0.48\textwidth]{contnoise25}

 \capt{The spectrum at a signal-to-noise level of 25, with the empirically determined continuum (as shown in fig. \ref{fig:noiselow}) subtracted.}
\label{fig:contnoise25}
\end{figure}

\begin{figure}[h!]

 \centering
  \includegraphics[trim=2cm 0cm 1.5cm 1.5cm, clip=true,width=0.48\textwidth]{contnoise05}

 \capt{This figure shows the spectrum in fig. \ref{fig:noisehigh}, at a signal-to-noise of 5, with the continuum subtracted.}
\label{fig:contnoise05}
\end{figure}\vskip-20pt
\end{comment}

The secondary step -- the denoising to determine the number of features -- is shown in figure \ref{fig:fclow}, for the continuum subtracted spectrum shown in \ref{fig:contnoiselow}. Note how the denoising artefacts (termed ringing) in figure \ref{fig:denoise} are less present, and as such are not mis-counted as features. In the noisier example (figure \ref{fig:contnoisehigh}) the denoising once again fails to detect any features and returns a null spectrum (for this particular noise realisation and FDR threshold of 4.55\% allowed false detections), and this would lead to the spectrum being discarded from our catalogue as unlikely to yield an accurate redshift estimate.

\begin{figure*}[h!]

 \centering
  \includegraphics[trim=2cm 0cm 1.5cm 1.5cm, clip=true,width=0.9\textwidth]{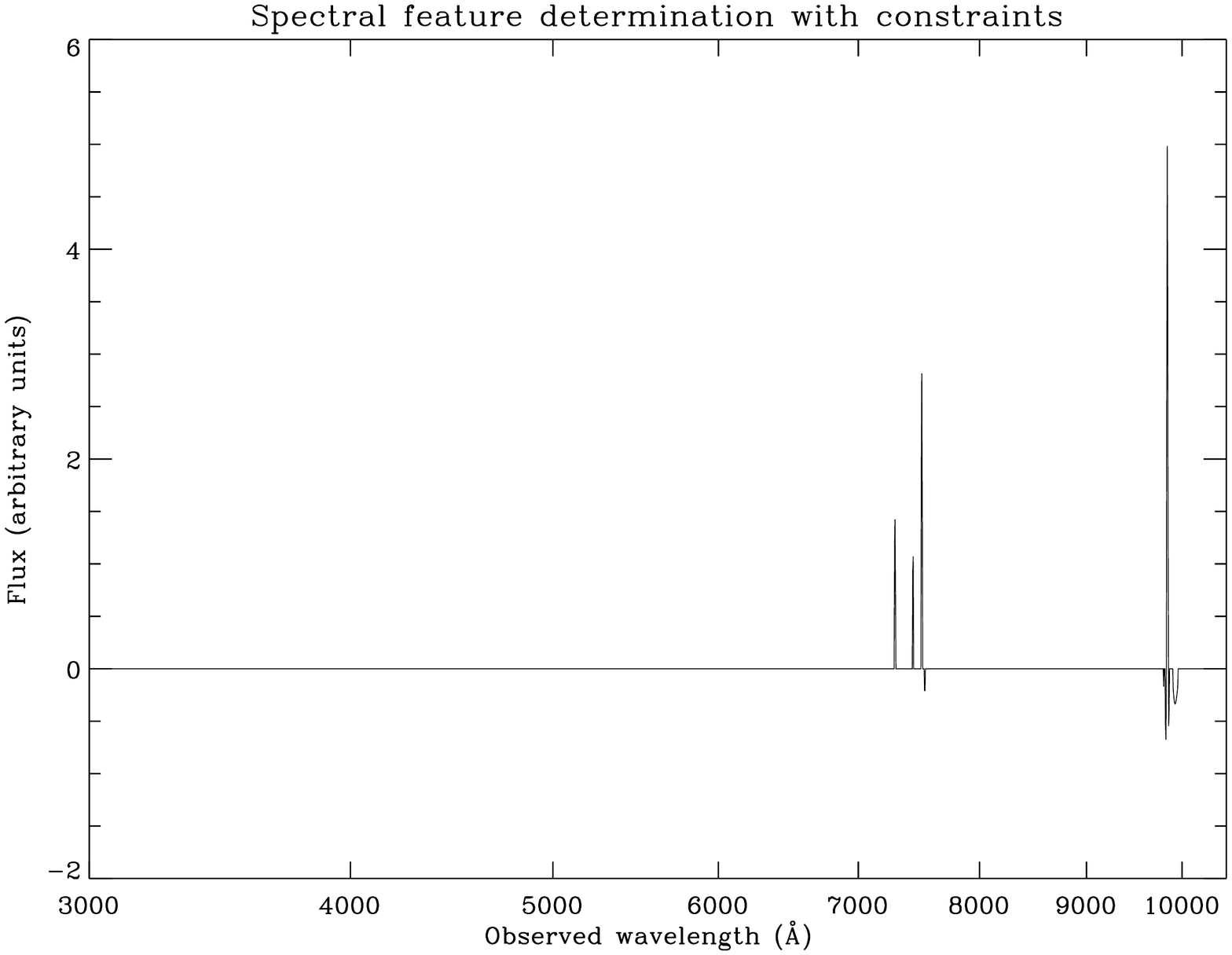}\vspace{-25pt}

 \capt{This figure shows the result of denoising the positive and negative sections (shown together) of the spectrum shown in figure \ref{fig:contnoiselow} with positivity and negativity constraints respectively. Note the reduced ringing, which leads to a more representative result with respect to the number of true features. Once again the FDR denoising of our noisier example (figure \ref{fig:contnoisehigh}) yields a null spectrum (not shown), and would thus result in the discarding of this spectrum from the redshift analysis.}
\label{fig:fclow}
\end{figure*}

\subsection{Redshift Estimation}\label{sec:zest}

For our method, we followed the PCA procedure as described in \sect{sec:pca} on a set of noise-free template spectra to obtain eigentemplates, of which we kept only the first $n$ principal components such that they comprised 99\% of the total eigenvalue weight, which in our case resulted in the retention of 20 eigentemplates. We continuum-subtracted our test spectra as described in section \sect{sec:modelling}. Since the white-Gaussian noise on a spectrum will in principle be uncorrelated with the eigentemplates, we chose to use the noisy galaxy spectra in the cross-correlation. This ensured that we preserved all the line information in the spectrum, rather than discarding some of the signal through denoising, and hence we were not probing potential systematics of the denoising simultaneously with the redshift estimation.

However, when dealing with the pixel-dependent noise, it is the denoised spectra that must be used in the cross-correlation, since very noisy pixels in proximity to less noisy pixels will produce features that strongly resemble lines, and would thus be highly correlated with features in the eigentemplates (independent of the redshift of the spectra involved) if not denoised. For example, an error-curve peaking strongly at \numprint{7600}\,\textup{\AA}, may frequently produce features at this wavelength in the noisy spectra that strongly resemble lines. The effect of this false feature is to bias the cross-correlations such that large features in the templates (for example H$_{\alpha}$) consistently match up to this false line, independent of the true redshift of the spectrum, resulting in redshift estimates that consistently favour an incorrect redshift. In this example many spectra would be biased to have an estimated redshift of 0.158, irrespective of their true redshift values. As such we must use the denoised versions of the spectra with non-stationary noise for the cross-correlations. However, the redshift estimation will thus incur any potential systematics of the denoising procedure itself, this is explored further in \sect{sec:realistic_errors}.

Clearly, at low SNR, some of these cross-correlations will produce inaccurate results due to many features becoming lost in the noise. Higher SNR is not a guarantee of a successful redshift estimate; it is possible that line confusion, a lack of features, or poor representation in the basis of eigentemplates will result in a catastrophic failure.

A simple, but effective, criterion for the selection of galaxy spectra that will be likely to yield an accurate redshift estimate can be developed by considering the number of significant line features (either absorption or emission) present in the spectrum. For a spectrum containing many prominent features, it should be very easy to determine the redshift via cross-correlation with a representative set of eigentemplates. In cases where only one prominent feature is present, for example, we expect that the cross-correlation function will have several local maxima, each occurring when the location of the line feature in the test spectrum aligns with any of the features present in the (shifted) template spectrum.  A similar effect would be expected for a spectrum with many -- but not particularly prominent -- features, obscured by noise. In such cases, it will not generally be possible to obtain a correct redshift unless we have more information about that feature or the spectrum (for example identifying the continuum shape/using photometric data which would help in identifying the colour of the galaxy; redder being indicative - but not definitively -  of higher redshift), and/or we make an assumption that the most prominent feature is a specific standard line (for example, H$_{\alpha}$). There is also the possibility that the dominant feature in the spectrum is a noise feature (this could be the case for multiple features if the spectrum is very noisy), in which case it will be impossible to estimate the redshift correctly.

With an increasing number of detected features, and a high degree of certainty that they are not the result of noise contamination, it should become clear that the redshift estimate obtained for such a test spectrum becomes progressively more reliable.

A question arises as to quite how many features are sufficient to distinguish reliable redshifts from those which are not reliable and we wish to discard. Through empirical tests, we have chosen 6 features in total as the criterion by which we decide the reliability of the redshift estimate of a test spectrum in our catalogue.

With this in mind, we use the denoising procedure described in \sect{sec:modelling} on the continuum-subtracted spectrum and identify the number of features present in the denoised spectrum via a simple feature-counting algorithm\footnote{Algorithm adapted from `peaks.pro', available from: \url{http://astro.berkeley.edu/~johnjohn/idlprocs/peaks.pro}}. We then partition the catalogue in two: a cleaned catalogue comprised of noisy spectra for which denoising presents 6 or more features, where we keep the redshift determination as likely to be accurate; and a discarded catalogue with spectra only possessing 5 features or fewer upon denoising, where the redshift estimates are deemed to be unreliable.

Features are considered to be `peaks' anywhere where the derivative of the spectrum changes from positive to negative (maxima), but only in the spectrum's positive domain; this means that, for example, a Gaussian-like function with two maxima (a line-doublet), would count as \emph{two} features. Employing this method alone would ignore absorption features; to overcome this we denoise and feature-count the positive and negative halves of the spectrum separately, independently detecting both emission and absorption features. %This is in part why the number of required features is high; some features, particularly spectral breaks such as the \numprint{4000}\,\textup{\AA} break, have the potential to lie in both the negative and positive halves of the spectrum (after continuum subtraction). Such features contribute twice to the total number of detected features, being counted both in the negative half and the positive half. This potential double-counting of features may mean that a spectrum with 6 detected features only corresponds to 5 features with a physical origin.

At low SNR there is a trade-off between relaxing the FDR threshold to pick up more features -- or indeed any features -- and imposing a stricter threshold to prevent the detection of spurious lines. Recall that the FDR parameter constrains the average ratio of false detections to total detections. Therefore, for an FDR parameter of $\alpha = 0.05$, for example, we allow on average one false feature for every 20 features detected; i.e. an average ratio of 19 true features to 1 false feature. In very noisy data, it might not be possible to achieve this statistical accuracy, and therefore no features will be identified by the algorithm.

It follows that even if 6 features can be obtained from the denoising of the spectrum, some of them may still be spurious, and this could lead to an erroneous redshift estimate from cross-correlation (particularly if the spurious line is dominant, with this strongly biasing the cross-correlation) and false-positive contamination of our retained data. However, as noted, a maximum for this false line contamination is set by the FDR threshold, $\alpha$. In addition, the spectra that possess fewer than 6 features may provide redshift estimates that would otherwise be reliable; the criterion chosen leads them to be discarded. There exists this necessary trade-off between the fraction of catastrophic failures in the resulting redshift catalogue and the amount of data that is discarded.

\section{Experimental Results}\label{sec:results}
In order to test our algorithm, we investigate the effect of the choice of the FDR parameter on the rate of catastrophic failures and the fraction of retained data in the cleaned catalogue at different SNR values. We use the simulated data described in \sect{sec:mock}, and apply the Darth Fader algorithm over multiple FDR thresholds, keeping the signal-to-noise constant; and again over catalogues with different SNRs, keeping the FDR threshold constant. Lastly we apply Darth Fader to a uniformly mixed SNR catalogue with pixel-dependent Gaussian noise with SNR ranging from 1 to 20, utilising a range of values for the FDR threshold.

We define the retention $\mathcal{R}$; catastrophic failure rates before cleaning, $\mathcal{F}$, and after cleaning, $\mathcal{F}_{c}$; and capture rate $\mathcal{C}$ of the sample to be:
\begin{eqnarray}
\ \ \ \ \ \ \mathcal{R} &=& \frac{\mathcal{T}_{c}}{\mathcal{T}} \times 100\% \: ,\label{compdef}\\
\ \ \ \ \ \ \mathcal{F}_{(c)} &=& \Bigg( 1-\frac{\mathcal{U}_{(c)}}{\mathcal{T}_{(c)}} \Bigg) \times 100\% \: ,\label{sucdef}\\
\ \ \ \ \ \ \mathcal{C} &=& \frac{\mathcal{U}_{c}}{\mathcal{U}} \times 100\% \: ,\label{capturedef}
\end{eqnarray}
where $\mathcal{T}$ and $\mathcal{T}_{c}$ respectively denote the total number of galaxies in the sample (before cleaning) and the retained number of galaxies in the sample after cleaning (the number that satisfy the feature-counting criterion). Similarly, $\mathcal{U}$ and $\mathcal{U}_{c}$, respectively denote the number of successful redshift estimates in the sample before and after cleaning. In equation \eqref{sucdef}, the brackets denote the option of calculating the catastrophic failure rate before cleaning (ignoring the subscripts) or the catastrophic failure rate after cleaning (inserting the subscript c everywhere shown). The number of successes after cleaning, $\mathcal{U}_{c}$, cannot be greater than $\mathcal{U}$, hence the capture rate represents the proportion of correct estimates available before cleaning that are retained post-cleaning.

\begin{figure*}[h]
\centering
        \begin{subfigure}[h!]{0.49\textwidth}
                %\centering
                \includegraphics[trim=1.0cm 6.0cm 1.0cm 5.5cm, clip=true,width=\textwidth]{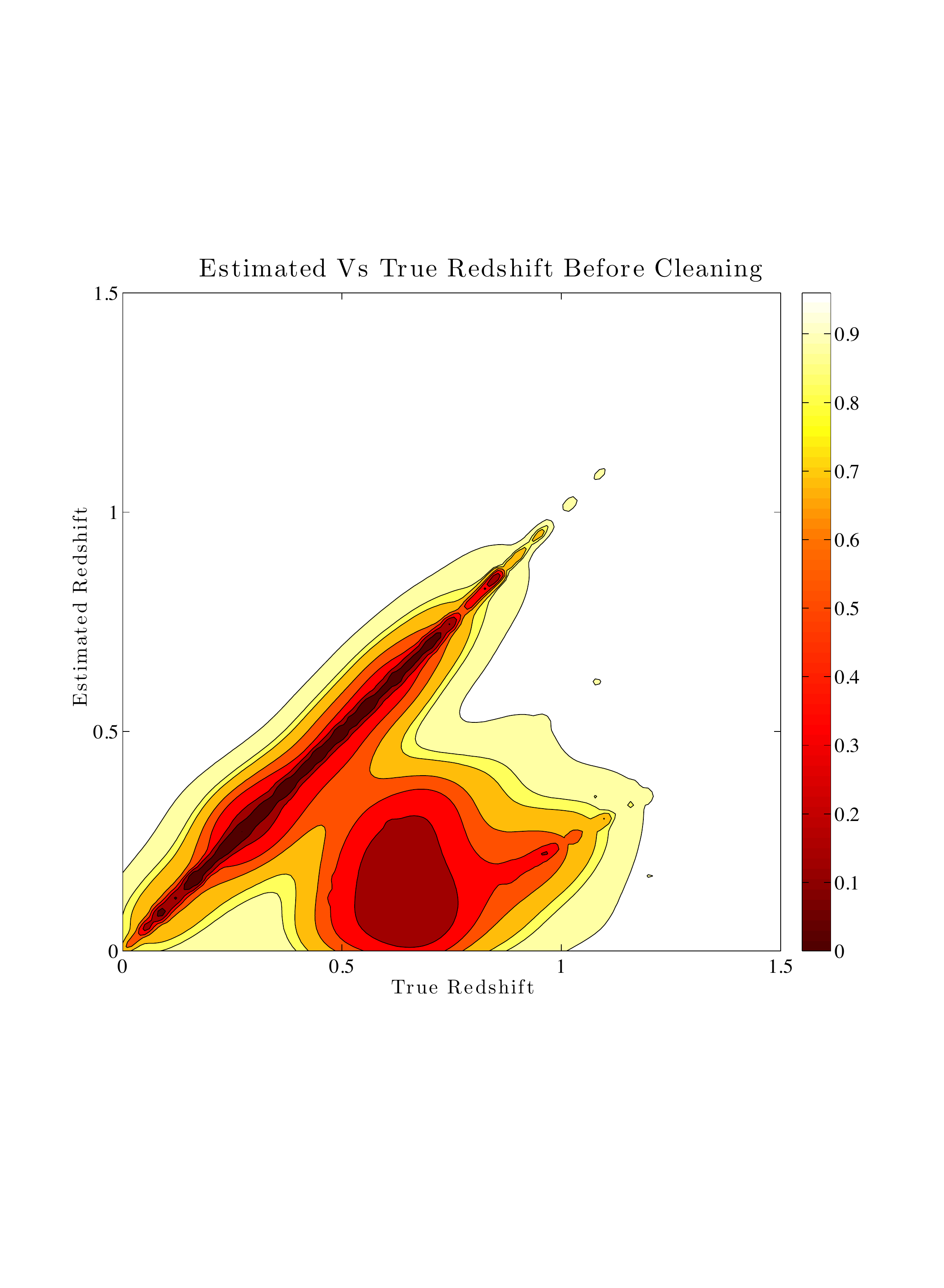}\vspace{-7pt}
                \caption{Before cleaning.}
                \label{fig:bc}
        \end{subfigure}\vspace{-4pt}
        ~
      \begin{subfigure}[h!]{0.49\textwidth}
               % \centering
                \includegraphics[trim=1.0cm 6.0cm 1.0cm 5.5cm, clip=true,width=\textwidth]{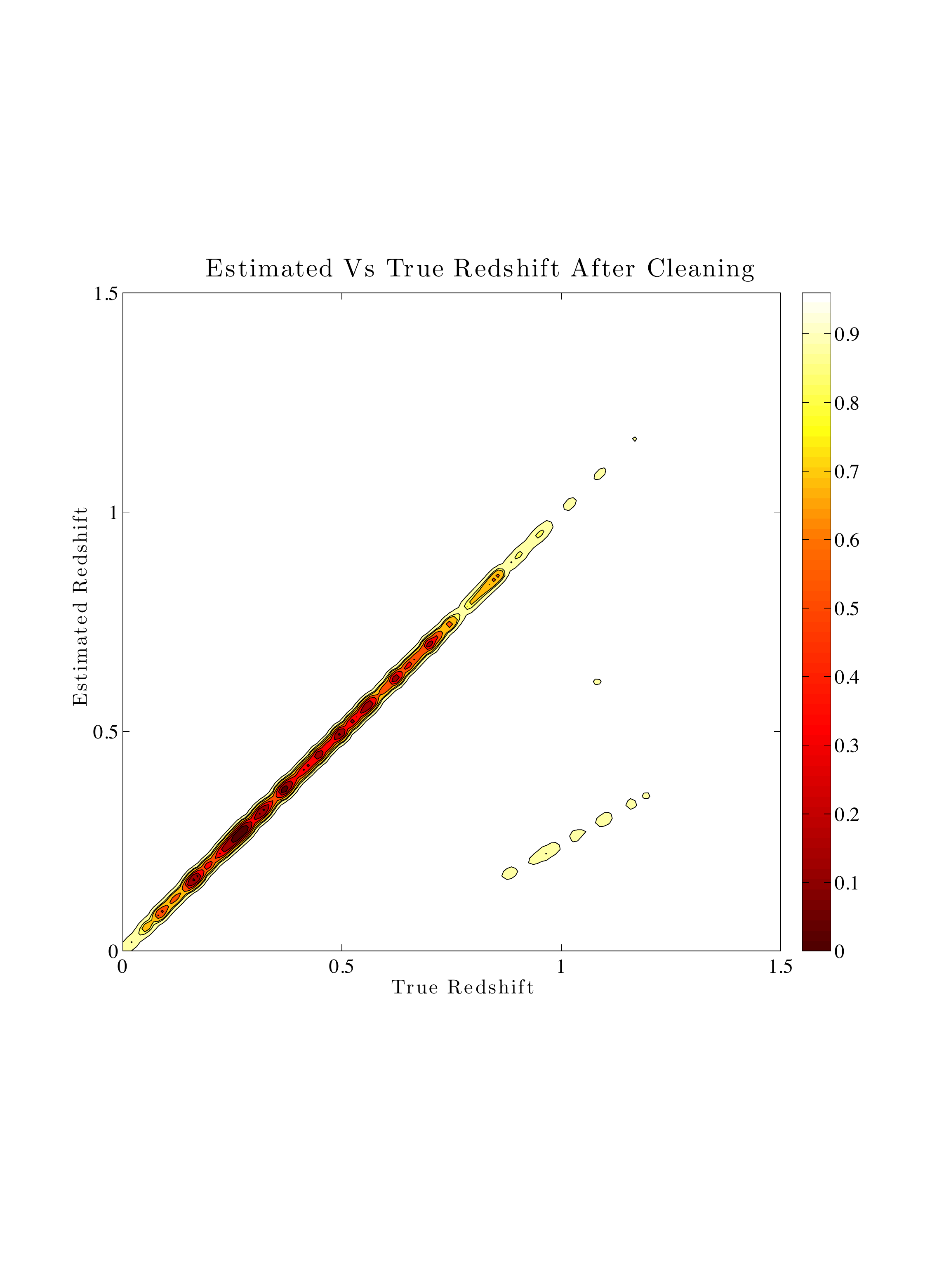}\vspace{-7pt}
                \caption{After cleaning.}
                \label{fig:ac}
              \end{subfigure}
        \caption{A contour plot to show the effect on redshift estimation before and after cleaning a catalogue which is at a signal-to-noise of 2.0, and cleaned with an FDR threshold of 4.55\% allowed false detections. Contours indicate the fraction of points enclosed within them. Figure \ref{fig:bc} depicts the results before cleaning, and \ref{fig:ac}, after; just under two thirds of all the estimated redshifts lie on the diagonal (and are thus correct) before cleaning being applied. Clearly outliers still exist after cleaning of the catalogue (off-diagonal), where the redshift estimation has failed, but as it can be seen, these are very few, and the result has a high certainty, with 94.9\% of the estimates being correct. The capture rate for this catalogue and at this FDR threshold is 76.2\%.}\label{fig:banda}
\end{figure*}

We present the result of cleaning the catalogue using an FDR threshold of $\alpha = 4.55\%$ on a catalogue of spectra with an SNR of 2.0 in figure \ref{fig:banda}. The two panels compare the distribution of redshift estimates before and after cleaning of the catalogue using the feature-counting criterion. A clear improvement is seen when cleaning is applied: the fraction of catastrophic failures in the catalogue is reduced from $34.5\%$ before cleaning to $5.1\%$ after cleaning. In addition, we have retained 76.2\% of the galaxies which yielded a correct redshift estimate before cleaning (the capture rate), with the retained catalogue comprising 52.6\% of the total number of galaxies in the test catalogue. 

Prior to cleaning there clearly exist two components to the redshift estimates, a strong square diagonal (the x=y line where the redshift estimates are likely correct) and a cloud of misidentifications (with a small, non-square, diagonal component) where the estimated redshifts are generally underestimates of the true redshift. It is important to note that failures at this point are due to the standard cross-correlation, with non-square diagonal components often being indicative of line confusion (for example between H$_{\alpha}$ and [O$\,$III]).

This represents a snapshot of how the Darth Fader algorithm works: we can \emph{blindly} isolate a correct subset of galaxies, ensuring a good coverage of the correct data available in the pre-cleaned catalogue, and we can -- by choosing an appropriate FDR threshold -- guarantee that the resultant catalogue contains a very low catastrophic failure rate. Though not implemented in Darth Fader, the data rejected could be subject to further analysis, using additional information (e.g. photometry) and alternative methodology to determine the redshifts of the galaxies.

\begin{figure*}[h!]

 \centering
  \includegraphics[trim=2cm 0cm 1.32cm 1cm, clip=true,width=0.9\textwidth]{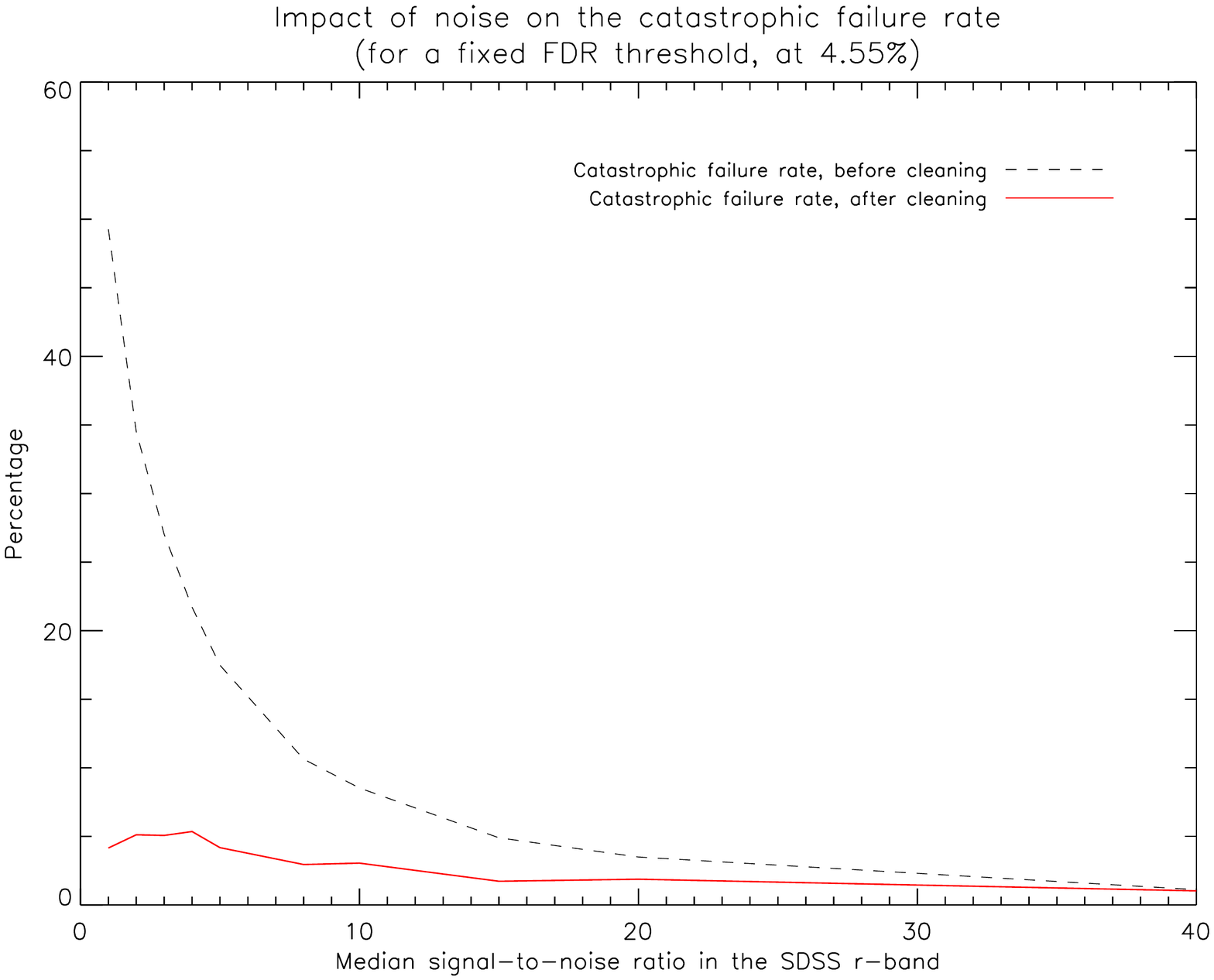}\vspace{-25pt}

 \capt{This figure illustrates how Darth Fader improves the catastrophic failure rate of the redshift estimates of the test catalogue at different signal-to-noise values (flat white-Gaussian noise) for a fixed FDR threshold of 4.55\% allowed false detections. Note the marked improvement in the SNR range 1.0-10.0 where catastrophic failure rates are reduced by up to 40\%. For this choice of $\alpha$, the catastrophic failure rate is always found to be $\lesssim$ 5\% after cleaning, for SNR values $\geq$ 1. Our catastrophic failure rate after cleaning at an SNR of 1 is similar to the rate for an SNR value of 15 without cleaning. The catastrophic failure rate before cleaning (dashed line) represents the theoretical minimum amount of data that must be discarded for perfect catalogue cleaning.} 
\label{fig:success}
\end{figure*}

Figure \ref{fig:success} shows the catastrophic failure rate of Darth Fader before and after catalogue cleaning for a fixed FDR threshold of $\alpha = 4.55\%$, as a function of median SNR in the r-band. At high SNR ($\sim$ 20), the standard cross-correlation method yields a low catastrophic failure rate, and cleaning yields little improvement. At an SNR of 10, however, the standard cross-correlation method experiences a progressive increase in the catastrophic failure rate, approaching 50\% at an SNR of 1; in contrast our method can maintain a low catastrophic failure rate ($\lesssim$ 5\%) for SNR levels of $\geq$ 1.

An important point to note is that the catastrophic failure rate before cleaning ($\mathcal{F}$) represents a theoretical minimum amount of data that \emph{must} be discarded with a perfect catalogue cleaning method (where $\mathcal{F}_{c}$ and $\mathcal{C}$ would be 0 and 100\% respectively); thus the theoretical maximum retention is given by 100\% - $\mathcal{F}$. In practice the amount discarded is usually greater (since we inevitably discard galaxies that would otherwise yield correct redshifts), but it can also be less than this if a more relaxed threshold is used, necessarily incurring false positive contamination in the retained data set.

Using an FDR threshold of 4.55\% allowed false detections, a catalogue of SNR = 2 has a catastrophic failure rate of 34.5\% before cleaning, thus our maximum expected retention in a perfect catalogue cleaning should only be 65.5\%, with our actual retention at that FDR threshold being 52.6\%. It should therefore not be surprising that at the lower end of signal-to-noise the expected retention values for a cleaned catalogue are (necessarily) low; however this can still represent a large proportion of the \emph{correct} data available. The recovery of these data still represents a net gain when compared to a total loss of these data, particularly when these recovered data can be known to be highly accurate.

At higher SNR, the impact of cleaning is reduced because denoising does not reveal more, significantly useful, diagnostic information: the number of features present in the noisy spectrum will more frequently already meet the selection criterion before cleaning, and thus cleaning the catalogue removes fewer spectra. To ensure a similarly low catastrophic failure rate in low SNR data would require a stricter FDR threshold to be used, and therefore would result in more data being discarded. 

To demonstrate the effect of the choice of FDR threshold on the catastrophic failure rate, the retention and the capture rate in the very low SNR regime, we test Darth Fader on two fixed low SNR catalogues of median SNR values of 1.0 and 2.0 with flat white-Gaussian noise, and one mixed SNR catalogue consisting of spectra with pixel-dependent Gaussian noise (see \ref{sec:realistic_errors}) with a uniform distribution of median SNR values between 1 and 20.

\begin{figure*}[h!]

 \centering
  \includegraphics[trim=1cm 0cm 1.0cm 0cm, clip=true,width=0.9\textwidth]{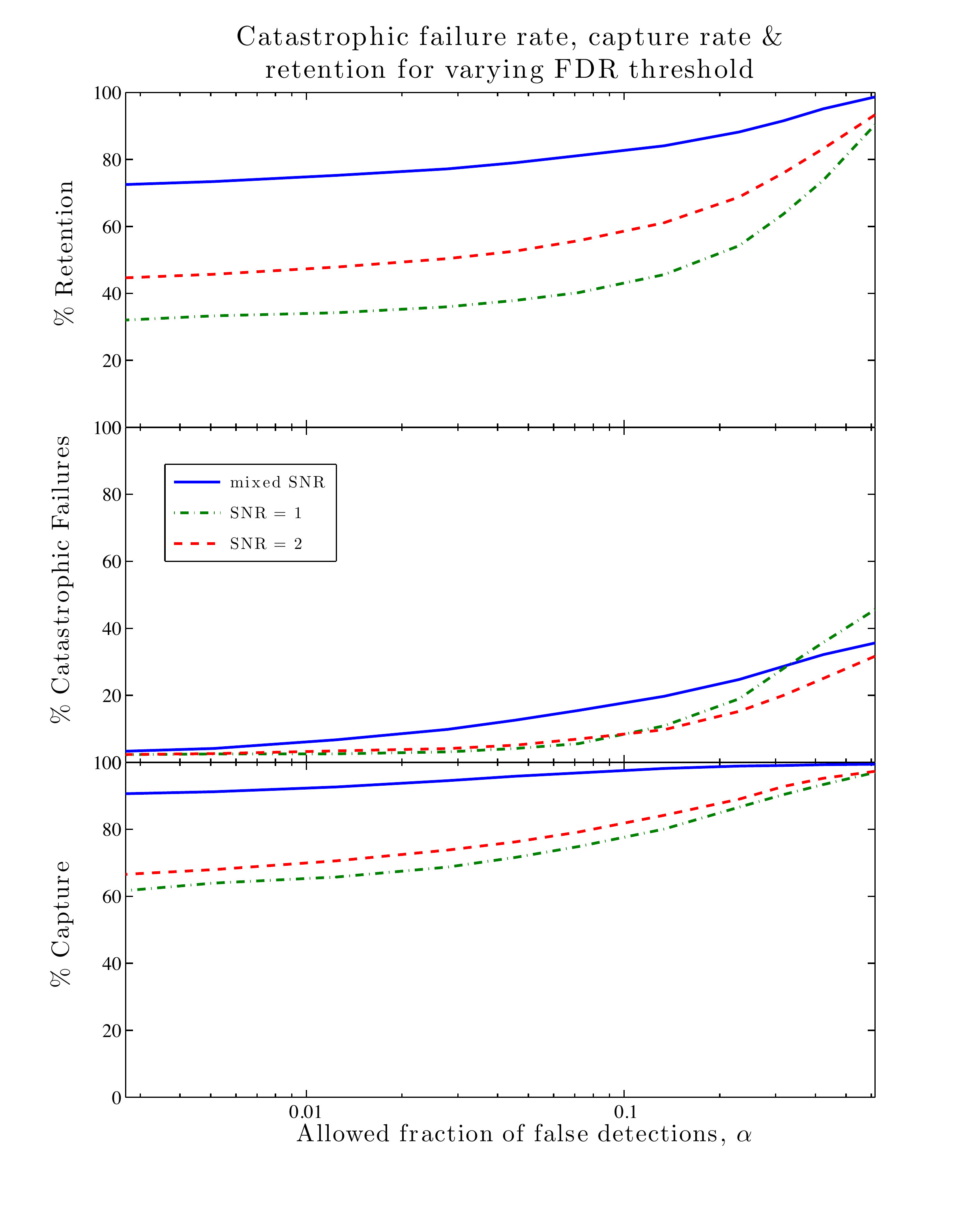}\vspace{-25pt}

 \capt{This figure illustrates the effect of the choice of FDR threshold on the catastrophic failure rate after cleaning, the retention and the capture rate on catalogues with a fixed SNR of 1.0 and 2.0 with flat noise; and on a mixed catalogue with a uniform distribution in SNR between 1 and 20, with pixel-dependent noise. Note the greater sacrifices required both in retention and capture rate in order to obtain the same catastrophic failure rate at an SNR of 1.0 compared to 2.0. Note also that we are able to obtain a 5.1\% failure rate in our redshift estimates for the cleaned catalogue, a retention of 52.6\%, and a capture rate of 76.2\% with the catalogue at an SNR of 2 at an FDR threshold of 4.55\%.}
\label{fig:varfdrlog}
\vspace{40pt}\end{figure*}

Figure \ref{fig:varfdrlog} clearly demonstrates the tradeoff that exists between the catastrophic failure rate after cleaning and the capture rate. Relaxing the threshold (i.e. increasing $\alpha$) improves both the retention and the capture rate by detecting more features in the spectra, more of which are likely to be false features rather than true ones, and thereby increasing the number of spectra accepted under the feature-counting criterion, but at a cost to the catastrophic failure rate since more erroneous spectra will also be accepted. A more conservative approach leads the FDR denoising to remove more real features, with the guarantee that very few of the remaining features will be false detections. This leads to a general decrease in both the retention and the capture rate since fewer spectra will exhibit the required number of features after denoising, with the benefit of this being a decrease in the catastrophic failure rate.

Notice also in figure \ref{fig:varfdrlog} that beyond a certain point the catastrophic failure rate saturates for the spectra with white-Gaussian noise (and shows little improvement for the mixed SNR catalogue with pixel-dependent noise), and stricter FDR thresholding (resulting in a smaller fraction of false detections) does not yield significant reductions in the rate of catastrophic failures; indeed this only serves to penalise both retention and the capture rate.

The results of the uniformly mixed and pixel-dependent SNR catalogue represent a step toward a more realistic view of what a real galaxy survey could look like. A real survey would not, however, have such a uniform distribution of SNR values, and would be skewed toward a greater population at lower SNR, with the actual distribution in signal-to-noise being specific to the type of survey and the instrument used.

\section{Application to Data}\label{sec:real_noise}

In the previous section, we demonstrated the robustness of the Darth Fader algorithm on simulations. Here we expand on the work in section \sect{sec:results} to show that our feature detection methods work well on real spectra from the SDSS archive.  

A full test on real SDSS data, for example, would not be practical since low SNR spectra are automatically discarded by magnitude cuts in the SDSS data processing pipeline and not available in their data products. Spectroscopic cuts for the main galaxy sample are taken at magnitudes in $r < 17.77$ (Petrosian) and the resulting spectra have a median SNR (per pixel) $> 4$ \citep{Strauss:2002}.

Real data differ from our simulations in a number of important ways: rare galaxy types/properties may exist within real data catalogues, and these may not necessarily be well encompassed by our simulations, and real data can often have more complex noise properties. It is therefore important to test whether our denoising methods, and feature-counting criterion, can be applied to real data.

\subsection{Realistic Pixel-dependent Noise}\label{sec:realistic_errors}

\begin{figure*}[h!]

 \centering
  \includegraphics[trim=2cm 0cm 1.5cm 1.5cm, clip=true,width=0.9\textwidth]{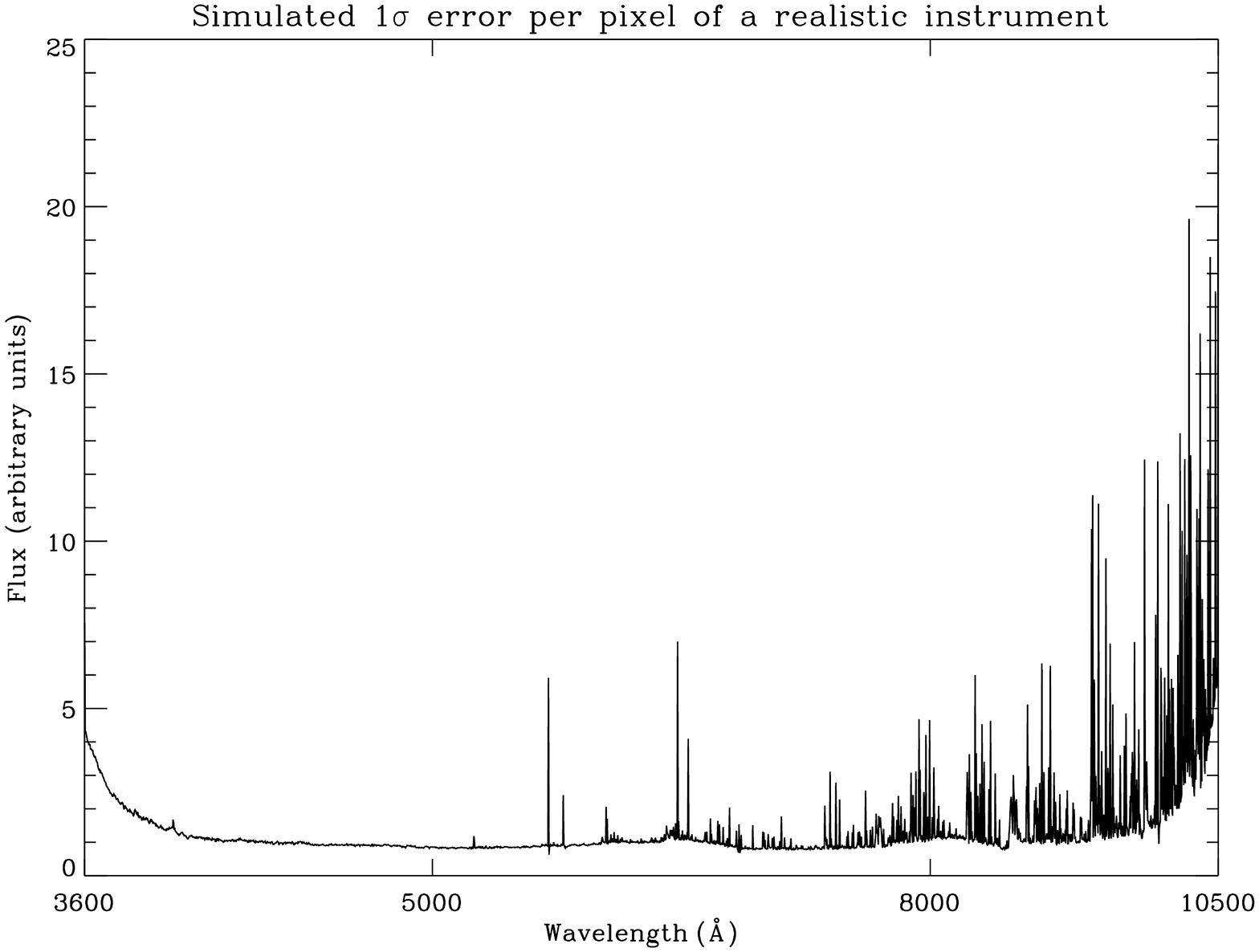}\vspace{-25pt}

 \capt{A realistic error-curve, where the resolution and binning are the same as for our mock catalogue, but with the wavelength range being slightly shorter, in order to be more proximal to the wavelength range of a realistic instrument. Gaussian noise is added to each pixel in our simulated data, with a standard deviation given by the value of the error-curve at that same pixel. }
\label{fig:noisecurve}
\end{figure*}

Real spectrographs have a sensitivity that varies -- sometimes quite strongly -- with wavelength or per pixel, primarily as a result of the sky brightness and instrumental efficiency. We simulate a realistic error-curve that spans the typical optical survey wavelength range, and in figure \ref{fig:noisecurve} we show the 1$\sigma$ error-curve per pixel used to mimic a realistic instrument. This is similar to what could be expected for an existing survey such as SDSS, or the forthcoming DESI spectroscopic survey \citep{Levi:2013}, (itself a merger of the BigBOSS \citep{Schlegel:2011} \& DESpec \citep{Abdalla:2012} spectroscopic surveys), as well as other projects involving multi-object spectrographs\footnote{These surveys are, however, expected to be at much higher resolution than our simulations.}. The Darth Fader algorithm can use the error-curve in the denoising step. Better accounting for the complex noise properties of the observed spectrum enhances the ability to discriminate between true features and those arising due to noise.

\begin{figure*}[h!]

 \centering
  \includegraphics[trim=2cm 0cm 1.5cm 1.5cm, clip=true,width=0.9\textwidth]{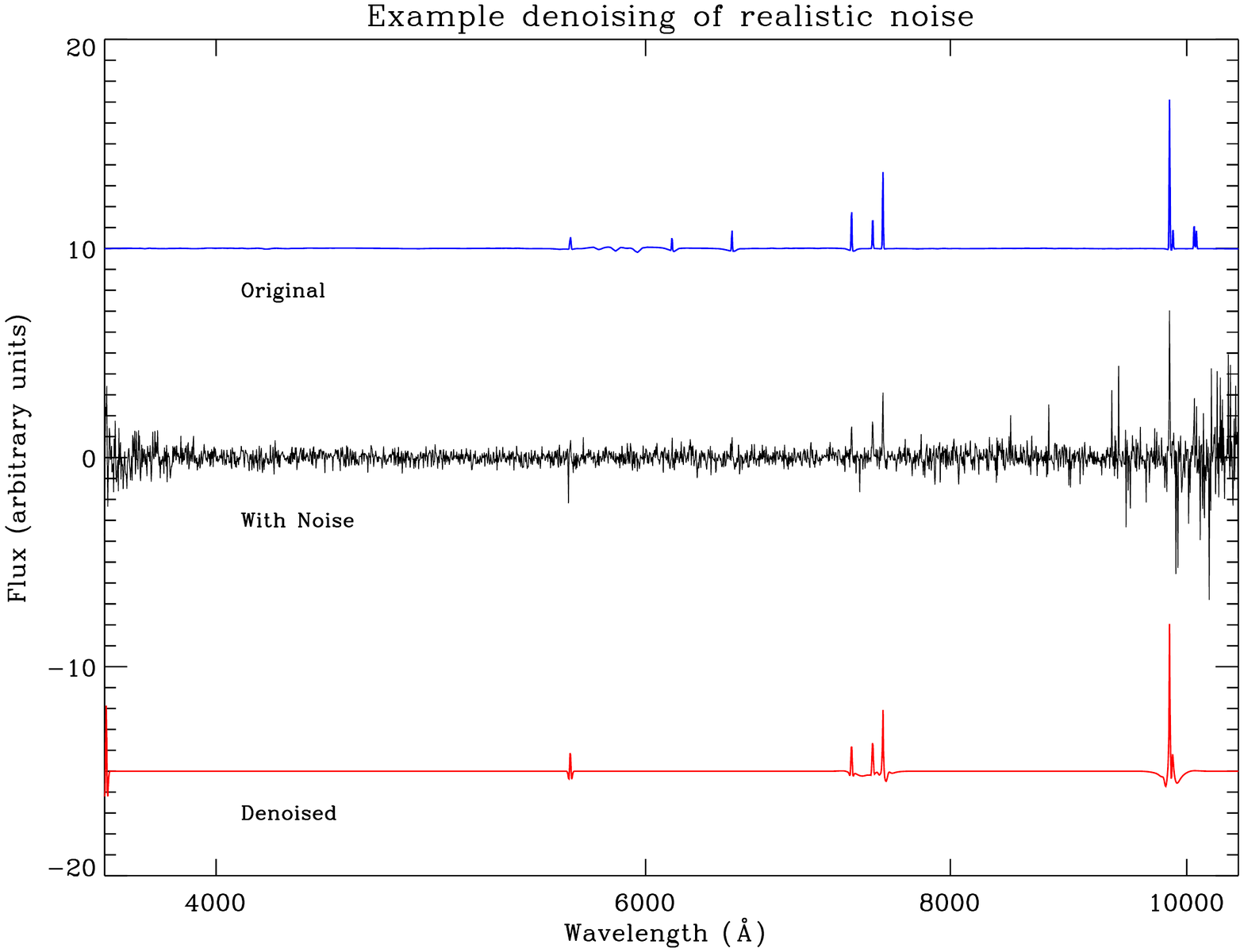}\vspace{-25pt}

 \capt{Denoising of test spectrum (c.f. figure \ref{fig:purespec}, continuum-subtracted) with pixel-dependent noise. Note how most of the main features are detected and how, for this particular noise realisation, no false detections are found in the complicated \& noisy long-wavelength region. We do incur a false detection at the very short-wavelength end of the spectrum. This is a systematic edge-effect resulting from a lack of information that would otherwise allow the algorithm to properly distinguish this as a noise feature.}
\label{fig:realistic_deno}
\end{figure*}

Figure \ref{fig:realistic_deno} shows a continuum-subtracted spectrum from our catalogue, truncated to match the wavelength range of the error-curve (and hence our simulated instrument), before and after the addition of the wavelength-dependent noise of figure \ref{fig:noisecurve}. We also plot the spectrum after denoising (again with an FDR threshold of 4.55\% allowed false detections) with the Darth Fader algorithm when supplied with the error-curve in figure \ref{fig:noisecurve}. This spectrum has a median SNR of 5 in the r-band at this particular redshift, ($z=1.5$). However, for the same noise level, this SNR would vary between 3 and 5 according to redshift, as a result of different continuum levels within the boundaries of the r-band.

\begin{figure*}[h!]

 \centering
  \includegraphics[trim=2cm 0cm 1.5cm 1.2cm, clip=true,width=0.9\textwidth]{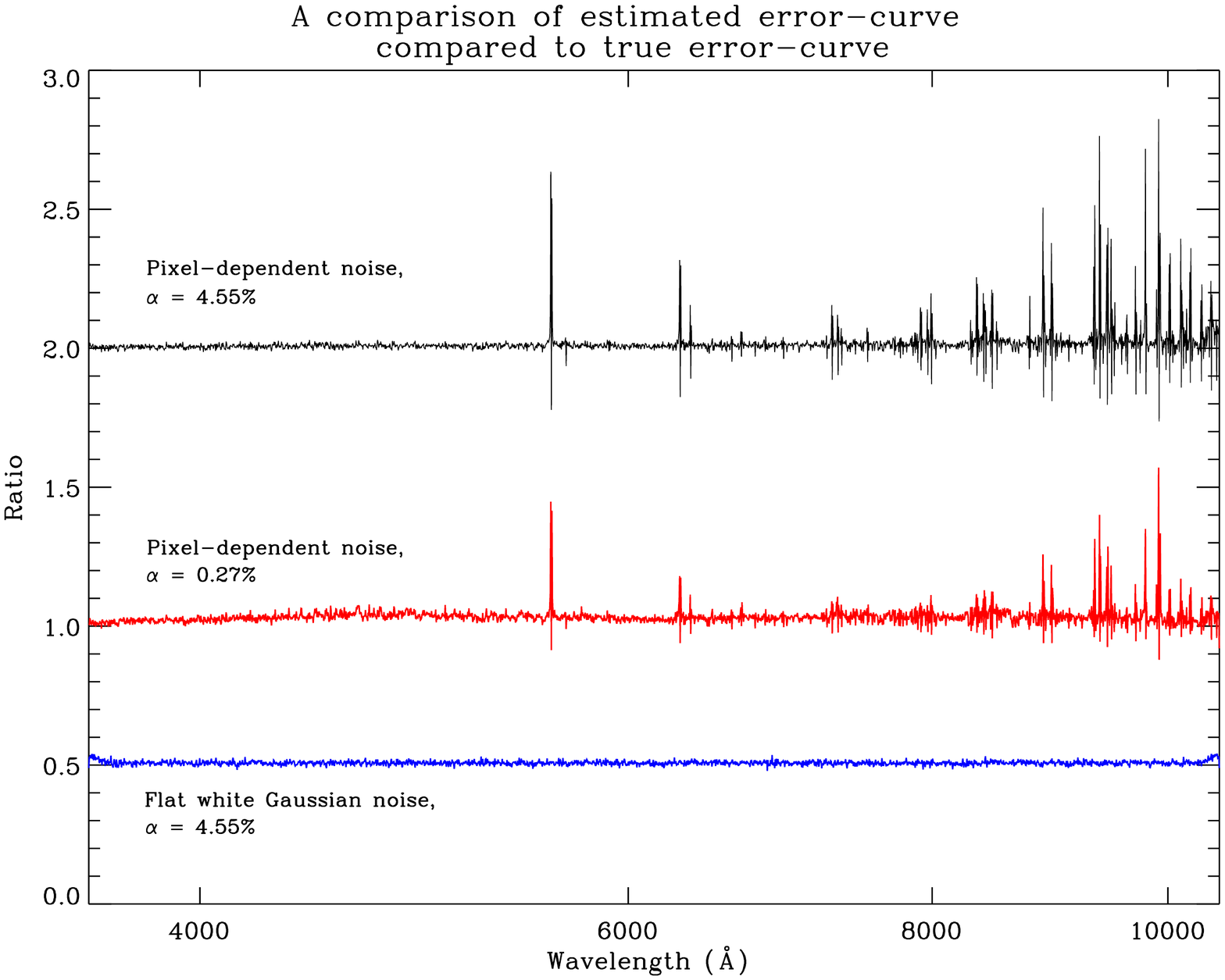}\vspace{-25pt}

 \capt{In this figure we plot the ratio of the true error-curve with respect to the derived error-curve from the rms error per pixel on the difference between the original input spectrum and the denoised spectrum for both flat noise and pixel-dependent noise. The lower curve (blue) has been shifted down (by 0.5) for clarity, and the upper curve (black), has also been shifted up (by 1.0) for clarity. Note the minor systematic edge effects on the denoising of white-Gaussian (flat) noise. Clearly the complex noise region has an marked systematic effect on the denoising, with rapidly changing noise regions experiencing both over and under estimates in the noise strength. This systematic effect is dependent upon the FDR threshold chosen, with thresholding that is less strict (upper curve) being more prone than stricter thresholding (middle curve).}
\label{fig:noisecurve_comp}
\end{figure*}

To test the effectiveness and robustness of the denoising, we use the same test spectrum as in fig. \ref{fig:realistic_deno}, and apply $10,000$ random (wrap-around) shifts in order to randomise the location of the principal features. For each shifted spectrum, pixel-dependent Gaussian noise is added as before, and at the same level. We then perform a denoising on each spectrum, and compute the residual with the input noisy spectrum. The RMS residual gives an estimate of the noise with its statistical distribution -- if the denoising has been effective -- matching the input error-curve. The randomised shifting of the spectrum allows us to determine the effectiveness of the denoising independently of the locations of the true features, and removes any cumulative biasing arising from features being undetected after denoising, or denoising artefacts. We do however expect a biasing at the edges of the spectrum at both the long and short-wavelength ends, due to a lack of information `beyond' the edge limiting the ability to correctly characterise local features at the edge as either signal or noise. In figure \ref{fig:noisecurve_comp}, we show the ratio of the noise standard deviation over the input error-curve as a function of wavelength, for both the pixel-dependent noise in figure \ref{fig:noisecurve}, at FDR parameters of $\alpha = 4.55\%$ and $\alpha = 0.27\%$, and for flat white Gaussian noise ($\alpha = 4.55\%$). The noise standard deviation has been computed from the $10,000$ residuals described above.

As can be seen in figure \ref{fig:noisecurve_comp}, the addition and subsequent denoising of flat noise behaves as one would expect; small deviations about a flat line situated at y=1 (shifted down in the figure for clarity). Minor artefacts are present at the edges due border effects specific to the wavelet transform, and features occasionally straddling the edges of the spectrum. The more complicated noise proves to be a considerably more difficult task than the flat noise, and clearly has some persistent residual features after denoising, particularly in the longer wavelength range where the error-curve is most complex. This discrepancy is due to the denoising not fully accounting for the rapidly changing noise from one pixel to the next.

Clearly this will impact on feature detection, resulting in a greater number spurious detections particularly at longer wavelengths. Increasing the FDR parameter, $\alpha$, does provide significant improvement in the efficacy of the denoising (as shown by the middle curve). It may be possible to further ameliorate these systematic effects with a more optimal wavelet choice (should one exist), or by assigning a weight to each pixel to counterbalance the effect of the denoising not fully accounting for the complex noise properties. Additionally, figure \ref{fig:varfdrlog} (solid blue line) shows that a stricter FDR thresholding is already effective in mitigating these systematic effects of the denoising.

We can conclude therefore that Darth Fader provides effective and robust denoising, regardless of whether the noise is stationary or wavelength dependent, provided that a choice of FDR parameter is made such that it is appropriate to the type of noise present.

\subsection{Feature Extraction in Real Data}\label{sec:real_data}

The number of features we can detect stems in part from the resolution of the spectra: if the resolution is poor, there is more uncertainty in the precise wavelength of the spectral lines, making it easier to confuse lines because they are slightly smeared out. Another, more important concern, is the potential blending of doublets or other lines in close proximity in wavelength, such as [N$\,$II] with H$_{\alpha}$. These localised groupings of lines provide powerful constraints on the galaxy redshift since the wavelength gap between the two lines in a doublet or close pair is often sufficient to conclusively isolate which emission lines are present, and hence deduce the redshift. Poorer resolution will often result in blending of such features, limiting the number of detectable features as well as broadening the possible location of the feature. Hence poor resolution impacts both the number of features through blending, and the detected locations of the features in wavelength due to coarser pixelisation of the data.

This can reduce the number of spectra meeting the feature-counting criterion in poorer resolution spectra, however this might be mitigated by considering a larger wavelength range for the spectra: provided features exist in this extended range, more features can be found to counteract the loss of feature detections and precision as a consequence of the poorer resolution. It is for this reason that our simulated spectra cover a larger wavelength range than SDSS currently does (however DESI is expected to have a similar wavelength range). In reality this trade-off is a minor consideration, however, since the practicalities of instrument design are the limiting factor for the wavelength range of spectra in real surveys. Our simulated spectra are at moderate resolution; instruments such as SDSS offer substantially higher resolution spectra.

In order to show the broader applicability of Darth Fader to real spectra potentially at higher resolution and covering a narrower wavelength range, we take three SDSS galaxy spectra an emission line galaxy, ELG, a luminous red galaxy, LRG, and a `typical' galaxy) and their respective 1$\sigma$ error-curves, as fiducial type galaxies that well represent the SDSS galaxy catalogue.\footnote{These three example galaxies can be found at: \url{ http://www.sdss.org/gallery/gal_spectra.html} with Plate ID: 312, MJD: 51689 and Fibre IDs: 220 (LRG), 255 (Typical), \& 529 (ELG).}

These spectra have a resolution $R \, \big( = \sfrac{\lambda}{\Delta\lambda}\big)$ significantly higher than that of our simulations, namely $R\sim$ \numprint{1845} compared to $R\sim$ 850, and as such features are better separated. The r-band SNR for these galaxies is quoted to be 9.2 for the ELG, 9.3 for the LRG, and 15.0 for the typical galaxy, respectively. 

In denoising these spectra we use an FDR threshold of $\alpha = 0.27\%$ as motivated by the discussion in \sect{sec:realistic_errors} and the results in figure \ref{fig:varfdrlog}. We apply a positivity (and `negativity') constraint, as before, to denoise the positive and negative sections of each spectrum independently, and recombine them to form the final denoised spectrum. The procedure uses the same positivity constraint, once on denoising the spectrum, and once on denoising the reverse-signed spectrum -- this is entirely equivalent to denoising once with a positivity constrain, and again with a `negativity' constraint.

\begin{figure*}[h!]

 \centering
  \includegraphics[trim=2cm 0cm 1.5cm 1.5cm, clip=true,width=0.9\textwidth]{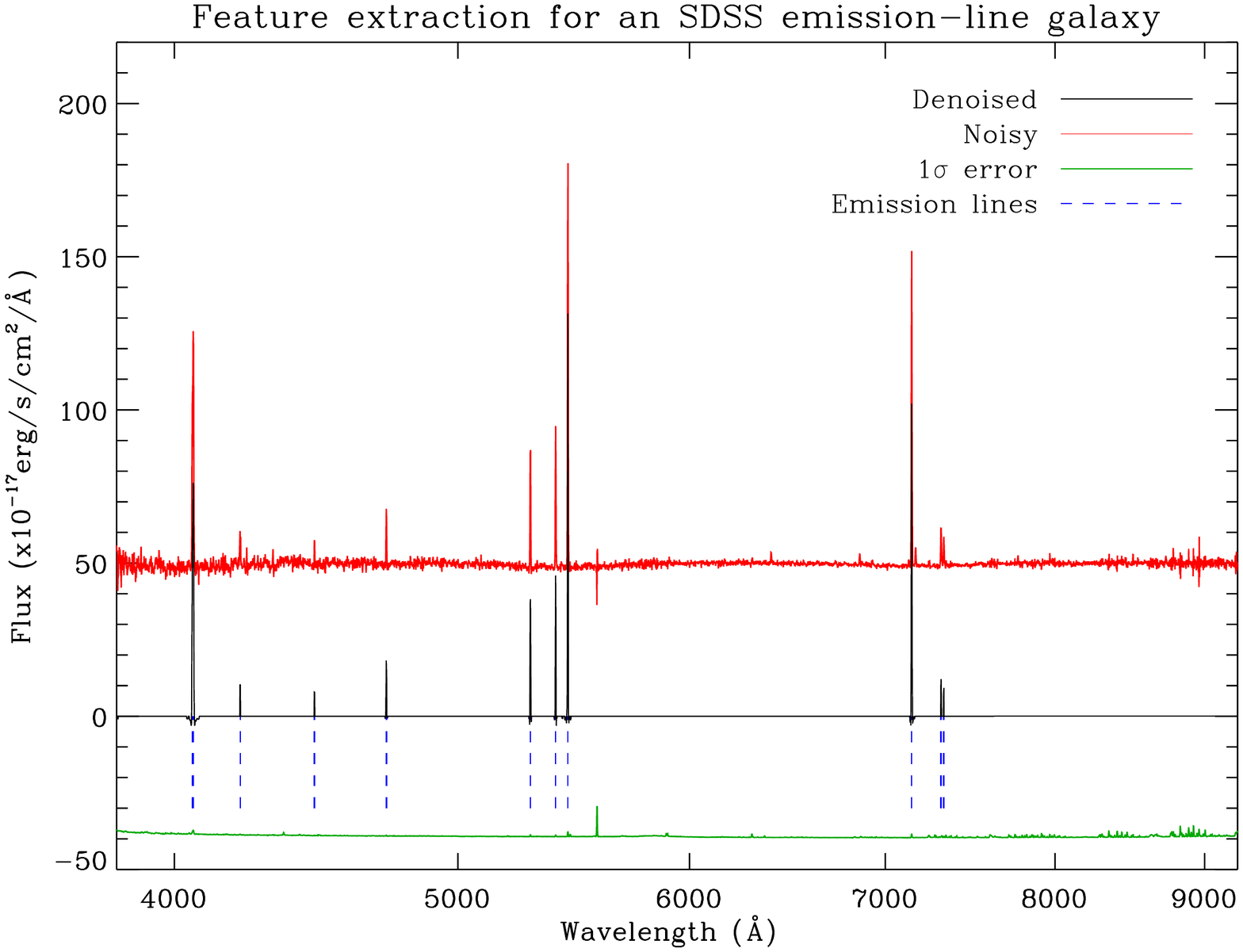}\vspace{-25pt}

 \capt{Denoising and feature extraction for an SDSS ELG. The noisy spectrum (red) has been shifted up, and the error-curve (green) shifted down, for clarity. The vertical dashed lines (blue) indicate the locations of detected features that correspond to true emission features. The FDR denoising and feature extraction clearly pinpoints all of the major features without any difficulty. The three largest lines are, from left to right, the [O$\,$II] doublet,  [O$\,$III] and H$_{\alpha}$.}
\label{fig:sdss_elg}
\end{figure*}

In figure \ref{fig:sdss_elg}, we show the continuum-subtracted spectrum, the FDR denoised spectrum, and the line features we detect for the emission-line galaxy. We also plot the 1$\sigma$ error-curve, which we assume as Gaussian. 

This ELG spectrum has many strong features, so it is not surprising that the FDR denoising detects most of them. We do however miss one very weak emission feature that is comparable to the noise, at $\sim $ 7,800 \textup{\AA}. It should also be noted that the potential line-like features arising from the noise, namely at $\sim $ 5,600 \textup{\AA} and again at $\sim $ 8,950 \textup{\AA} are completely ignored by the FDR denoising since, by supplying the error-curve, these features are correctly identified as likely arising from noise rather than signal.

\begin{figure*}[h!]

 \centering
  \includegraphics[trim=2cm 0cm 1.5cm 1.5cm, clip=true,width=0.9\textwidth]{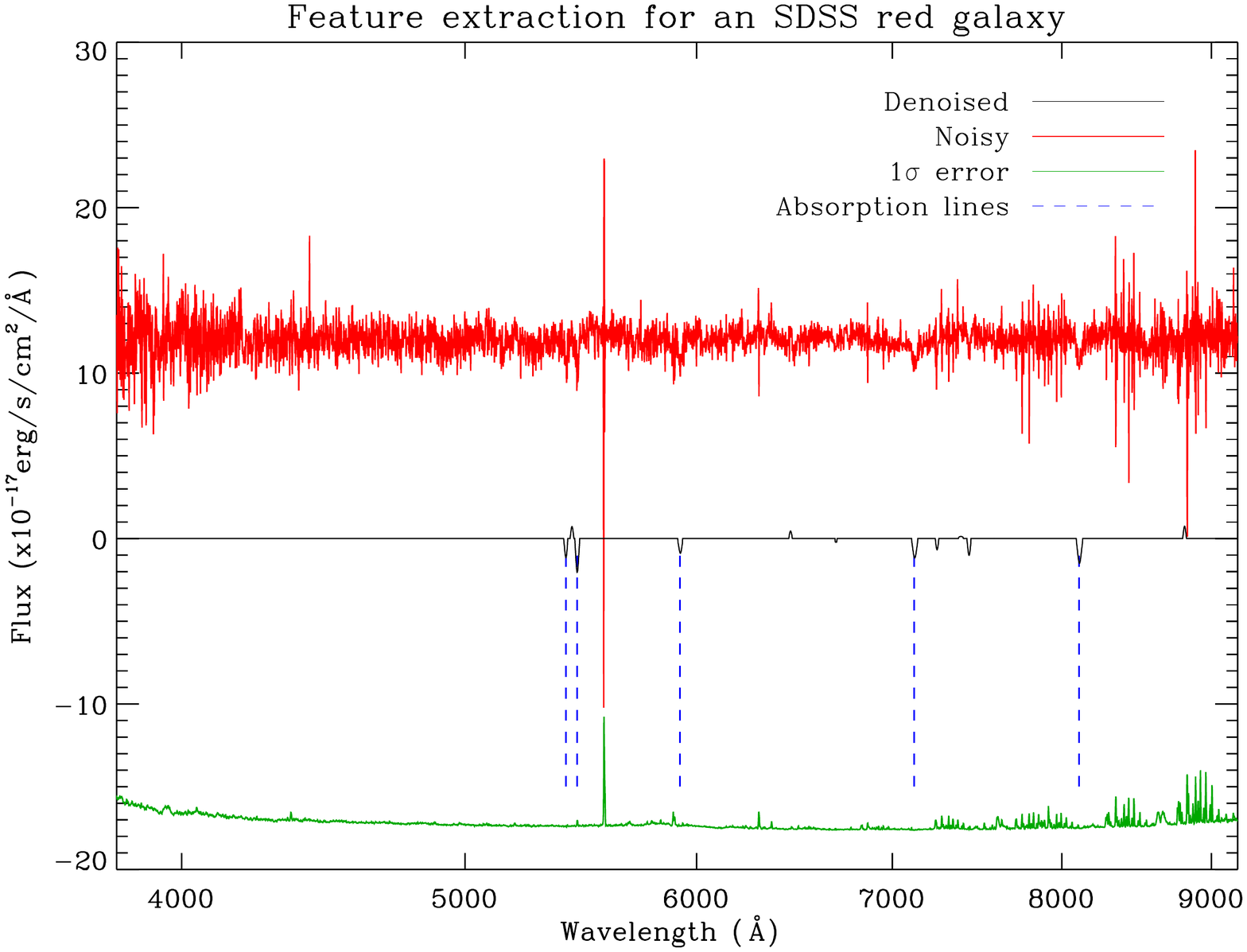}\vspace{-25pt}

 \capt{Denoising and feature extraction for an SDSS LRG. The absorption lines from left to right are CaII (H and K), G-band, MgI and NaI.\protect\footnotemark}
\label{fig:sdss_lrg}
\end{figure*}

Feature detection in the LRG spectrum (fig. \ref{fig:sdss_lrg}) presents a more difficult challenge. Despite the signal-to-noise ratio in the r-band being of similar value to the ELG, the widths of the features compared to the noise (i.e. the signal-to-noise values on the lines) are much smaller. We successfully detect five absorption features, despite them not being particularly prominent. We detect further spurious, smaller, features that cannot be associated with any common lines, and are likely minor artefacts from denoising.

\begin{figure*}[h!]

 \centering
  \includegraphics[trim=2cm 0cm 1.5cm 1.5cm, clip=true,width=0.9\textwidth]{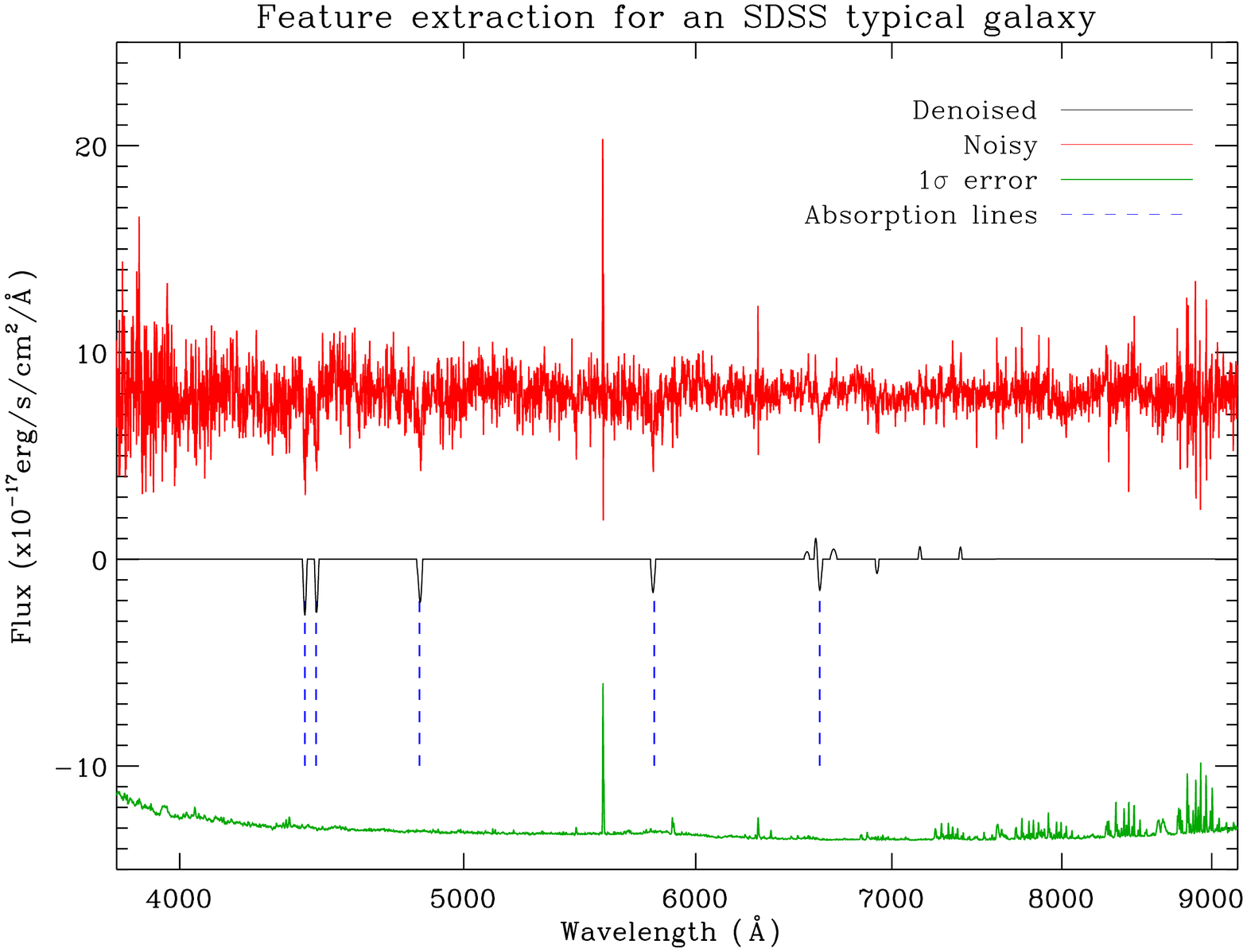}\vspace{-25pt}

 \capt{Denoising and feature extraction for an SDSS typical galaxy. This spectrum is similar to that of the LRG, the highlighted absorption lines being the same as previously.}
\label{fig:sdss_typ}
\end{figure*}

The results for the typical galaxy are similar to those of the LRG (fig. \ref{fig:sdss_typ}). In this case, we again detect all five of the absorption features, in addition we obtain some unidentifiable spurious features.

For each of the galaxy types shown here we can detect at least six features, though not all of them are true detections, nor do we require them to be necessarily true detections. Though we only consider Gaussian noise here, the tools used in Darth Fader are in principle not limited to purely Gaussian errors, and can be utilised with different types of errors (in particular Poisson, Gaussian + Poisson, multiplicative and correlated errors), and provided that denoising can be done appropriately the impact on the Darth Fader method will be minimal.

\section{Conclusions}\label{sec:conclusions}

As we have shown, Darth Fader is a powerful tool for the improvement of redshift estimation without any \emph{a priori} knowledge of galactic composition, type or morphology.

We can successfully make an estimate of the continuum without needing to model the spectra, and we can confidently make use of data at signal-to-noise levels that were previously beyond the reach of other techniques. This is achieved by denoising the data with an appropriately chosen false detection rate threshold and implementing a simple feature-counting criterion, resulting in very low catastrophic failure rate for redshift estimates for a subset of galaxies in the catalogue.

This is the most useful aspect of Darth Fader -- it can be used as a flagging mechanism to extract what is likely to be good data for redshift estimation from what is likely to yield an inaccurate redshift estimate, with a good level of confidence. Even at signal-to-noise levels as low as 2.0 in the r-band, we can retain 52.6\% of the data, and contained within this subset we can obtain 76.2\% of all the potentially correct redshift estimates that were initially available, resulting in a highly confident subsample where 94.9\% of the redshift estimates are reliable. This cleaning therefore has applications in large surveys, such as the upcoming Euclid survey \citep{euclid,euclid2}, which requires a spectroscopic redshift catalogue with very few catastrophic failures. 

Darth Fader represents a potential greater reach of spectroscopic surveys in terms of depth, since the faintest (and thus noisiest) galaxies in a survey -- those at the detection limit of the instrument -- will be tend to be those at higher redshifts. Currently, lower signal to noise spectra tend to be discarded, or to yield highly unreliable redshift estimates. Darth Fader allows for the inclusion of a substantial subset of these otherwise-discarded galaxies.

Darth Fader demonstrates that these current methods, with a blunt cut-off in the signal-to-noise/flux for what is considered to be informative data, can be significantly improved upon, and that improvements are available all the way down to very low signal-to-noise levels. The levels of retention presented in this paper may seem moderate, however, for such low signal-to-noise data they can only be expected to be so, as redshift estimation necessarily fails for a large fraction of spectra at these high noise levels. Darth Fader performs very well and can reliably extract the majority of the likely correct data that is available in the low SNR regime. It is for this reason that for our method the capture rate is a more useful diagnostic than the retention, since it shows that for the available informative data contained within the uninformative data, Darth Fader can isolate and extract the majority of it in a blind and fully automated manner, resulting in catalogues where we can have a very high degree of certainty that the redshifts are correct. Indeed, for all the SNR levels we test, Darth Fader is always able to capture at least 60\% of the available data that we know to be correct, with this minimum rising to $70\%$ with more standard choices of FDR parameter values ($\sim$ 4\%). Hence -- even when overall retention values appear moderate -- the low rate of catastrophic failures together with the high proportion of good data available that is retained, represent a substantial gain when the alternative is to throw away the entire dataset with a blunt SNR cut because we cannot be certain as to which spectra are viable or not.

\footnotetext{Note: G-band absorption is strictly not an absorption \emph{line}, but rather an aggregate absorption feature due to the presence of multiple lines arising from metals (mainly iron) in the numerous G-type stars present in the galaxy population. Also not to be confused with the SDSS photometric filter g-band.}

In addition, Darth Fader can deal with realistic noise providing we know the 1$\sigma$ error-curve, where we can reduce catastrophic failure rates of a mixed SNR catalogue from 22.7\% down to 3.3\% with a capture rate of 90.6\% with an appropriate choice of FDR parameter. The algorithm does a good job of capturing nearly all of the spectra for which we are able to obtain a correct redshift estimate, and simultaneously maintaining a low catastrophic failure rate. Furthermore, we have shown that the continuum subtraction and feature-identification methods used in Darth Fader are effective on spectra from the SDSS data archive. This provides a proof-of-concept of the applicability of the method to real data, though there may be scope for development in this area. An additional feature is that, depending on different needs, the FDR parameter can be adjusted to enhance the capture rate and retention at a small cost to the catastrophic failure rate, or vice versa.

Our analysis in this paper utilises catalogues which are simulated, and as such they may be simpler than catalogues of real data; they do however fully represent the expected various galactic morphological types, distribution and redshift properties. The noise properties we use in our simulations may also be less complicated than those in real data, however, non-stationary Gaussian noise (varying per pixel) is a good enough approximation to real data, and this has been shown with the competent denoising of the SDSS example spectra (figures \ref{fig:sdss_elg}, \ref{fig:sdss_lrg}, \ref{fig:sdss_typ}).

The simulations we use are of considerably lower resolution ($R\sim$ 850 compared to $R\sim$ \numprint{1845}) than would be expected for a modern day spectral survey, with SDSS resolution being over twice as high, and the forthcoming DESI survey \citep{Levi:2013} -- a merger of the BigBOSS and DESpec surveys \citep[][respectively]{Schlegel:2011,Abdalla:2012} -- expected to be higher still. The wavelength range of our simulated spectra (\numprint{3000}\,\textup{\AA} to \numprint{10500}\,\textup{\AA}) are slightly longer than would be expected for a realistic instrument (\numprint{3500} $\lesssim \lambda \lesssim$ \numprint{10000}), but given the poorer resolution in our simulations, it is justifiable to extend the range. These factors do not, however, prevent these catalogues from being realistic. Indeed we have shown that it is possible to detect the required number of features in a shorter wavelength range, such as that of SDSS. There is therefore great promise for the use of these techniques in future large scale structure surveys, for feature extraction, redshift determination, and photometric calibration.

Continuum removal with wavelets, when compared against elaborate modelling, may be seen as a comparatively na\"{i}ve method. However, there is no loss of generality in its usage in cross-correlation based redshift estimation methods, and it benefits from being a blind method requiring no prior knowledge of how galactic spectra arise. 

The wavelet-based continuum subtraction procedure used in Darth Fader is in principle not limited to galactic spectra, and preliminary tests suggest that it will prove useful for the continuum-modelling of the more structurally rich spectra of stars. Indeed, for any spectra whose components are easily modelled with the correct choice of wavelet, we expect our continuum subtraction method to work as demonstrated.

Although we only consider the numbers of features in this paper, the ability of the Darth Fader algorithm to detect likely true lines could readily be adapted to deal with feature \emph{identification}, in particular for spectra where noise levels are very high, and $K\sigma$ clipping would offer little advantage (since it has a tendency to clip signal as well as noise). This would make it possible to cross-check the position of standard lines which have been redshifted to match the estimated redshift of the galaxy, against the positions of the maxima in the FDR denoised spectrum (since these are the features considered important by FDR).

Darth Fader is clearly useful for both redshift estimation and empirical continuum estimation and will be made publicly available as part of the \textbf{iSAP}\footnote{\textbf{iSAP} package available at: \url{http://www.cosmostat.org/software.html}} suite of codes. The blind nature of our algorithm, together with the ability to handle realistic noise, show promise for its inclusion in future spectral survey pipelines and data analyses.

%-----------------------------------------------------------------------------------------------------------------------------------
%% Back matter
%-----------------------------------------------------------------------------------------------------------------------------------

%% Bibliography

\bibliographystyle{aa}
\clearpage

\balance

\begin{acknowledgements}
The authors would like to acknowledge the support provided by the European Research Council, through grant SparseAstro (ERC-228261), FBA acknowledges support from the Royal Society via an URF. We also wish to thank J\'{e}r\'{e}my Rapin \& Simon Beckouche for their helpful feedback and assistance in the data presentation. The authors would also like to thank the anonymous referee for his/her thoughts and contributions, which have helped to improve the quality of this paper.

\end{acknowledgements}

\bibliography{Darth_Fader_v1.1}

\begin{thebibliography}{40}
\expandafter\ifx\csname natexlab\endcsname\relax\def\natexlab#1{#1}\fi

\bibitem[{{Abdalla} {et~al.}(2012){Abdalla}, {Annis}, {Bacon}, {Bridle},
  {Castander}, {Colless}, {DePoy}, {Diehl}, {Eriksen}, {Flaugher}, {Frieman},
  {Gaztanaga}, {Hogan}, {Jouvel}, {Kent}, {Kirk}, {Kron}, {Kuhlmann}, {Lahav},
  {Lawrence}, {Lin}, {Marriner}, {Marshall}, {Mohr}, {Nichol}, {Sako},
  {Saunders}, {Soares-Santos}, {Thomas}, {Wechsler}, {West}, \&
  {Wu}}]{Abdalla:2012}
{Abdalla}, F., {Annis}, J., {Bacon}, D., {et~al.} 2012, ArXiv e-prints,
  arXiv:1209.2451

\bibitem[{{Aihara} {et~al.}(2011){Aihara}, {Allende Prieto}, {An}, {Anderson},
  {Aubourg}, {Balbinot}, {Beers}, {Berlind}, {Bickerton}, {Bizyaev}, {Blanton},
  {Bochanski}, {Bolton}, {Bovy}, {Brandt}, {Brinkmann}, {Brown}, {Brownstein},
  {Busca}, {Campbell}, {Carr}, {Chen}, {Chiappini}, {Comparat}, {Connolly},
  {Cortes}, {Croft}, {Cuesta}, {da Costa}, {Davenport}, {Dawson}, {Dhital},
  {Ealet}, {Ebelke}, {Edmondson}, {Eisenstein}, {Escoffier}, {Esposito},
  {Evans}, {Fan}, {Femen{\'{\i}}a Castell{\'a}}, {Font-Ribera}, {Frinchaboy},
  {Ge}, {Gillespie}, {Gilmore}, {Gonz{\'a}lez Hern{\'a}ndez}, {Gott}, {Gould},
  {Grebel}, {Gunn}, {Hamilton}, {Harding}, {Harris}, {Hawley}, {Hearty}, {Ho},
  {Hogg}, {Holtzman}, {Honscheid}, {Inada}, {Ivans}, {Jiang}, {Johnson},
  {Jordan}, {Jordan}, {Kazin}, {Kirkby}, {Klaene}, {Knapp}, {Kneib},
  {Kochanek}, {Koesterke}, {Kollmeier}, {Kron}, {Lampeitl}, {Lang}, {Le Goff},
  {Lee}, {Lin}, {Long}, {Loomis}, {Lucatello}, {Lundgren}, {Lupton}, {Ma},
  {MacDonald}, {Mahadevan}, {Maia}, {Makler}, {Malanushenko}, {Malanushenko},
  {Mandelbaum}, {Maraston}, {Margala}, {Masters}, {McBride}, {McGehee},
  {McGreer}, {M{\'e}nard}, {Miralda-Escud{\'e}}, {Morrison}, {Mullally},
  {Muna}, {Munn}, {Murayama}, {Myers}, {Naugle}, {Neto}, {Nguyen}, {Nichol},
  {O'Connell}, {Ogando}, {Olmstead}, {Oravetz}, {Padmanabhan},
  {Palanque-Delabrouille}, {Pan}, {Pandey}, {P{\^a}ris}, {Percival},
  {Petitjean}, {Pfaffenberger}, {Pforr}, {Phleps}, {Pichon}, {Pieri}, {Prada},
  {Price-Whelan}, {Raddick}, {Ramos}, {Reyl{\'e}}, {Rich}, {Richards}, {Rix},
  {Robin}, {Rocha-Pinto}, {Rockosi}, {Roe}, {Rollinde}, {Ross}, {Ross},
  {Rossetto}, {S{\'a}nchez}, {Sayres}, {Schlegel}, {Schlesinger}, {Schmidt},
  {Schneider}, {Sheldon}, {Shu}, {Simmerer}, {Simmons}, {Sivarani}, {Snedden},
  {Sobeck}, {Steinmetz}, {Strauss}, {Szalay}, {Tanaka}, {Thakar}, {Thomas},
  {Tinker}, {Tofflemire}, {Tojeiro}, {Tremonti}, {Vandenberg}, {Vargas
  Maga{\~n}a}, {Verde}, {Vogt}, {Wake}, {Wang}, {Weaver}, {Weinberg}, {White},
  {White}, {Yanny}, {Yasuda}, {Yeche}, \& {Zehavi}}]{Aihara:2011}
{Aihara}, H., {Allende Prieto}, C., {An}, D., {et~al.} 2011, Astrophysical
  Journal Supplement Series, 193, 29

\bibitem[{Benjamini \& Hochberg(1995)}]{Benjamini:1995}
Benjamini, Y. \& Hochberg, Y. 1995, J. R. Stat. Soc. B, 57, 289

\bibitem[{{Bolton} {et~al.}(2012){Bolton}, {Schlegel}, {Aubourg}, {Bailey},
  {Bhardwaj}, {Brownstein}, {Burles}, {Chen}, {Dawson}, {Eisenstein}, {Gunn},
  {Knapp}, {Loomis}, {Lupton}, {Maraston}, {Muna}, {Myers}, {Olmstead},
  {Padmanabhan}, {P{\^a}ris}, {Percival}, {Petitjean}, {Rockosi}, {Ross},
  {Schneider}, {Shu}, {Strauss}, {Thomas}, {Tremonti}, {Wake}, {Weaver}, \&
  {Wood-Vasey}}]{Bolton:2012}
{Bolton}, A.~S., {Schlegel}, D.~J., {Aubourg}, {\'E}., {et~al.} 2012,
  Astronomical Journal, 144, 144

\bibitem[{{Bruzual} \& {Charlot}(2003)}]{Bruzual:2003}
{Bruzual}, G. \& {Charlot}, S. 2003, Monthly Notices of the Royal Astronomical
  Society, 344, 1000

\bibitem[{{Capak}(2009)}]{Capak:2009}
{Capak}, P.~L. 2009, in American Astronomical Society Meeting Abstracts, Vol.
  214, American Astronomical Society Meeting Abstracts \#214, \#200.06

\bibitem[{{Costero} \& {Osterbrock}(1977)}]{Costero:1977}
{Costero}, R. \& {Osterbrock}, D.~E. 1977, Astrophysical Journal, 211, 675

\bibitem[{Fadili \& Starck(2009)}]{Fadili:2009}
Fadili, M.~J. \& Starck, J.-L. 2009, in Proceedings of the International
  Conference on Image Processing, ICIP 2009, 7-10 November 2009, Cairo, Egypt
  (IEEE), 1461--1464

\bibitem[{{Fligge} \& {Solanki}(1997)}]{Fligge:1997}
{Fligge}, M. \& {Solanki}, S.~K. 1997, Astronomy \& Astrophysics Supplement
  Series, 124, 579

\bibitem[{{Fukugita} {et~al.}(1996){Fukugita}, {Ichikawa}, {Gunn}, {Doi},
  {Shimasaku}, \& {Schneider}}]{Fukugita:1996}
{Fukugita}, M., {Ichikawa}, T., {Gunn}, J.~E., {et~al.} 1996, Astronomical
  Journal, 111, 1748

\bibitem[{{Garilli} {et~al.}(2010){Garilli}, {Fumana}, {Franzetti}, {Paioro},
  {Scodeggio}, {Le F{\`e}vre}, {Paltani}, \& {Scaramella}}]{Garilli:2010}
{Garilli}, B., {Fumana}, M., {Franzetti}, P., {et~al.} 2010, Publications of
  the Astronomical Society of the Pacific, 122, 827

\bibitem[{{Glazebrook} {et~al.}(1998){Glazebrook}, {Offer}, \&
  {Deeley}}]{Glazebrook:1998}
{Glazebrook}, K., {Offer}, A.~R., \& {Deeley}, K. 1998, Astrophysical Journal,
  492, 98

\bibitem[{{Hopkins} {et~al.}(2002){Hopkins}, {Miller}, {Connolly}, {Genovese},
  {Nichol}, \& {Wasserman}}]{Hopkins:2002}
{Hopkins}, A.~M., {Miller}, C.~J., {Connolly}, A.~J., {et~al.} 2002,
  Astrophysical Journal, 123, 1086

\bibitem[{{Ilbert} {et~al.}(2009){Ilbert}, {Capak}, {Salvato}, {Aussel},
  {McCracken}, {Sanders}, {Scoville}, {Kartaltepe}, {Arnouts}, {Le Floc'h},
  {Mobasher}, {Taniguchi}, {Lamareille}, {Leauthaud}, {Sasaki}, {Thompson},
  {Zamojski}, {Zamorani}, {Bardelli}, {Bolzonella}, {Bongiorno}, {Brusa},
  {Caputi}, {Carollo}, {Contini}, {Cook}, {Coppa}, {Cucciati}, {de la Torre},
  {de Ravel}, {Franzetti}, {Garilli}, {Hasinger}, {Iovino}, {Kampczyk},
  {Kneib}, {Knobel}, {Kovac}, {Le Borgne}, {Le Brun}, {F{\`e}vre}, {Lilly},
  {Looper}, {Maier}, {Mainieri}, {Mellier}, {Mignoli}, {Murayama}, {Pell{\`o}},
  {Peng}, {P{\'e}rez-Montero}, {Renzini}, {Ricciardelli}, {Schiminovich},
  {Scodeggio}, {Shioya}, {Silverman}, {Surace}, {Tanaka}, {Tasca}, {Tresse},
  {Vergani}, \& {Zucca}}]{Ilbert:2009}
{Ilbert}, O., {Capak}, P., {Salvato}, M., {et~al.} 2009, Astrophysical Journal,
  690, 1236

\bibitem[{{Jouvel} {et~al.}(2009){Jouvel}, {Kneib}, {Ilbert}, {Bernstein},
  {Arnouts}, {Dahlen}, {Ealet}, {Milliard}, {Aussel}, {Capak}, {Koekemoer}, {Le
  Brun}, {McCracken}, {Salvato}, \& {Scoville}}]{Jouvel:2009}
{Jouvel}, S., {Kneib}, J.-P., {Ilbert}, O., {et~al.} 2009, Astronomy and
  Astrophysics, 504, 359

\bibitem[{{Koski} \& {Osterbrock}(1976)}]{Koski:1976}
{Koski}, A.~T. \& {Osterbrock}, D.~E. 1976, Astrophysical Journal, Letters,
  203, L49

\bibitem[{{Kurtz} \& {Mink}(1998)}]{Kurtz:1998}
{Kurtz}, M.~J. \& {Mink}, D.~J. 1998, Publications of the Astronomical Society
  of the Pacific, 110, 934

\bibitem[{{Laureijs} {et~al.}(2011){Laureijs}, {Amiaux}, {Arduini},
  {Augu{\`e}res}, {Brinchmann}, {Cole}, {Cropper}, {Dabin}, {Duvet}, {Ealet},
  \& et~al.}]{euclid2}
{Laureijs}, R., {Amiaux}, J., {Arduini}, S., {et~al.} 2011, ArXiv e-prints

\bibitem[{{Le F{\`e}vre} {et~al.}(1995){Le F{\`e}vre}, {Crampton}, {Lilly},
  {Hammer}, \& {Tresse}}]{LeFevre:1995}
{Le F{\`e}vre}, O., {Crampton}, D., {Lilly}, S.~J., {Hammer}, F., \& {Tresse},
  L. 1995, Astrophysical Journal, 455, 60

\bibitem[{{Le F{\`e}vre} {et~al.}(2005){Le F{\`e}vre}, {Vettolani}, {Garilli},
  {Tresse}, {Bottini}, {Le Brun}, {Maccagni}, {Picat}, {Scaramella},
  {Scodeggio}, {Zanichelli}, {Adami}, {Arnaboldi}, {Arnouts}, {Bardelli},
  {Bolzonella}, {Cappi}, {Charlot}, {Ciliegi}, {Contini}, {Foucaud},
  {Franzetti}, {Gavignaud}, {Guzzo}, {Ilbert}, {Iovino}, {McCracken}, {Marano},
  {Marinoni}, {Mathez}, {Mazure}, {Meneux}, {Merighi}, {Paltani}, {Pell{\`o}},
  {Pollo}, {Pozzetti}, {Radovich}, {Zamorani}, {Zucca}, {Bondi}, {Bongiorno},
  {Busarello}, {Lamareille}, {Mellier}, {Merluzzi}, {Ripepi}, \&
  {Rizzo}}]{LeFevre:2005}
{Le F{\`e}vre}, O., {Vettolani}, G., {Garilli}, B., {et~al.} 2005, Astronomy \&
  Astrophysics, 439, 845

\bibitem[{{Levi} {et~al.}(2013){Levi}, {Bebek}, {Beers}, {Blum}, {Cahn},
  {Eisenstein}, {Flaugher}, {Honscheid}, {Kron}, {Lahav}, {McDonald}, {Roe},
  {Schlegel}, \& {representing the DESI collaboration}}]{Levi:2013}
{Levi}, M., {Bebek}, C., {Beers}, T., {et~al.} 2013, ArXiv e-prints

\bibitem[{{Lutz} {et~al.}(2008){Lutz}, {Schuh}, {Silvotti}, {Dreizler},
  {Green}, {Fontaine}, {Stahn}, {H{\"u}gelmeyer}, \& {Husser}}]{Lutz:2008}
{Lutz}, R., {Schuh}, S., {Silvotti}, R., {et~al.} 2008, in Astronomical Society
  of the Pacific Conference Series, Vol. 392, Hot Subdwarf Stars and Related
  Objects, ed. U.~{Heber}, C.~S. {Jeffery}, \& R.~{Napiwotzki}, 339

\bibitem[{{Miller} {et~al.}(2001){Miller}, {Genovese}, {Nichol}, {Wasserman},
  {Connolly}, {Reichart}, {Hopkins}, {Schneider}, \& {Moore}}]{Miller:2001}
{Miller}, C.~J., {Genovese}, C., {Nichol}, R.~C., {et~al.} 2001, Astrophysical
  Journal, 122, 3492

\bibitem[{{Panuzzo} {et~al.}(2007){Panuzzo}, {Vega}, {Bressan}, {Buson},
  {Clemens}, {Rampazzo}, {Silva}, {Vald\a'es}, {Granato}, \&
  {Danese}}]{Panuzzo:2007}
{Panuzzo}, P., {Vega}, O., {Bressan}, A., {et~al.} 2007, Astrophysical Journal,
  656, 206

\bibitem[{{Pires} {et~al.}(2006){Pires}, {Juin}, {Yvon}, {Moudden}, {Anthoine},
  \& {Pierpaoli}}]{Pires:2006}
{Pires}, S., {Juin}, J.~B., {Yvon}, D., {et~al.} 2006, Astronomy \&
  Astrophysics, 455, 741-755

\bibitem[{{Refregier} {et~al.}(2010){Refregier}, {Amara}, {Kitching}, {Rassat},
  {Scaramella}, {Weller}, \& {Euclid Imaging Consortium}}]{euclid}
{Refregier}, A., {Amara}, A., {Kitching}, T.~D., {et~al.} 2010, ArXiv:
  1001.0061

\bibitem[{{Schlegel} {et~al.}(2011){Schlegel}, {Abdalla}, {Abraham}, {Ahn},
  {Allende Prieto}, {Annis}, {Aubourg}, {Azzaro}, {Baltay}, {Baugh}, {Bebek},
  {Becerril}, {Blanton}, {Bolton}, {Bromley}, {Cahn}, {Carton},
  {Cervantes-Cota}, {Chu}, {Cortes}, {Dawson}, {Dey}, {Dickinson}, {Diehl},
  {Doel}, {Ealet}, {Edelstein}, {Eppelle}, {Escoffier}, {Evrard}, {Faccioli},
  {Frenk}, {Geha}, {Gerdes}, {Gondolo}, {Gonzalez-Arroyo}, {Grossan},
  {Heckman}, {Heetderks}, {Ho}, {Honscheid}, {Huterer}, {Ilbert}, {Ivans},
  {Jelinsky}, {Jing}, {Joyce}, {Kennedy}, {Kent}, {Kieda}, {Kim}, {Kim},
  {Kneib}, {Kong}, {Kosowsky}, {Krishnan}, {Lahav}, {Lampton}, {LeBohec}, {Le
  Brun}, {Levi}, {Li}, {Liang}, {Lim}, {Lin}, {Linder}, {Lorenzon}, {de la
  Macorra}, {Magneville}, {Malina}, {Marinoni}, {Martinez}, {Majewski},
  {Matheson}, {McCloskey}, {McDonald}, {McKay}, {McMahon}, {Menard},
  {Miralda-Escude}, {Modjaz}, {Montero-Dorta}, {Morales}, {Mostek}, {Newman},
  {Nichol}, {Nugent}, {Olsen}, {Padmanabhan}, {Palanque-Delabrouille}, {Park},
  {Peacock}, {Percival}, {Perlmutter}, {Peroux}, {Petitjean}, {Prada},
  {Prieto}, {Prochaska}, {Reil}, {Rockosi}, {Roe}, {Rollinde}, {Roodman},
  {Ross}, {Rudnick}, {Ruhlmann-Kleider}, {Sanchez}, {Sawyer}, {Schimd},
  {Schubnell}, {Scoccimaro}, {Seljak}, {Seo}, {Sheldon}, {Sholl},
  {Shulte-Ladbeck}, {Slosar}, {Smith}, {Smoot}, {Springer}, {Stril}, {Szalay},
  {Tao}, {Tarle}, {Taylor}, {Tilquin}, {Tinker}, {Valdes}, {Wang}, {Wang},
  {Weaver}, {Weinberg}, {White}, {Wood-Vasey}, {Yang}, {Yeche}, {Zakamska},
  {Zentner}, {Zhai}, \& {Zhang}}]{Schlegel:2011}
{Schlegel}, D., {Abdalla}, F., {Abraham}, T., {et~al.} 2011, ArXiv e-prints,
  arXiv:1106.1706

\bibitem[{{Smee} {et~al.}(2012){Smee}, {Gunn}, {Uomoto}, {Roe}, {Schlegel},
  {Rockosi}, {Carr}, {Leger}, {Dawson}, {Olmstead}, {Brinkmann}, {Owen},
  {Barkhouser}, {Honscheid}, {Harding}, {Long}, {Lupton}, {Loomis}, {Anderson},
  {Annis}, {Bernardi}, {Bhardwaj}, {Bizyaev}, {Bolton}, {Brewington}, {Briggs},
  {Burles}, {Burns}, {Castander}, {Connolly}, {Davenport}, {Ebelke}, {Epps},
  {Feldman}, {Friedman}, {Frieman}, {Heckman}, {Hull}, {Knapp}, {Lawrence},
  {Loveday}, {Mannery}, {Malanushenko}, {Malanushenko}, {Merrelli}, {Muna},
  {Newman}, {Nichol}, {Oravetz}, {Pan}, {Pope}, {Ricketts}, {Shelden},
  {Sandford}, {Siegmund}, {Simmons}, {Smith}, {Snedden}, {Schneider},
  {Strauss}, {SubbaRao}, {Tremonti}, {Waddell}, \& {York}}]{Smee:2012}
{Smee}, S., {Gunn}, J.~E., {Uomoto}, A., {et~al.} 2012, ArXiv e-prints,
  arXiv:1208.2233

\bibitem[{Starck {et~al.}(1995)Starck, Bijaoui, \& Murtagh}]{Starck:1995}
Starck, J.-L., Bijaoui, A., \& Murtagh, F. 1995, CVGIP: Graphical Models and
  Image Processing, 57, 420--431

\bibitem[{Starck {et~al.}(1996{\natexlab{a}})Starck, Claret, \&
  Siebenmorgen}]{Starck:1996_tech}
Starck, J.-L., Claret, A., \& Siebenmorgen, R. 1996{\natexlab{a}}, {ISOCAM}
  Data Calibration, Tech. rep., CEA

\bibitem[{{Starck} \& {Murtagh}(1994)}]{Starck:1994}
{Starck}, J.-L. \& {Murtagh}, F. 1994, Astronomy \& Astrophysics, 288, 342

\bibitem[{{Starck} \& {Murtagh}(2006)}]{Starck:2006}
{Starck}, J.-L. \& {Murtagh}, F. 2006, {Astronomical Image and Data Analysis}
  (Springer), 2nd edn.

\bibitem[{Starck {et~al.}(2010)Starck, Murtagh, \& Fadili}]{Starck:2010}
Starck, J.-L., Murtagh, F., \& Fadili, M. 2010, Sparse Image and Signal
  Processing (Cambridge University Press)

\bibitem[{Starck {et~al.}(1996{\natexlab{b}})Starck, Murtagh, Pirenne, \&
  Albrecht}]{Starck:1996}
Starck, J.-L., Murtagh, F., Pirenne, B., \& Albrecht, M. 1996{\natexlab{b}},
  Publications of the Astronomical Society of the Pacific, 108, 446--455

\bibitem[{{Stoughton} {et~al.}(2002){Stoughton}, {Lupton}, {Bernardi},
  {Blanton}, {Burles}, {Castander}, {Connolly}, {Eisenstein}, {Frieman},
  {Hennessy}, {Hindsley}, {Ivezi{\'c}}, {Kent}, {Kunszt}, {Lee}, {Meiksin},
  {Munn}, {Newberg}, {Nichol}, {Nicinski}, {Pier}, {Richards}, {Richmond},
  {Schlegel}, {Smith}, {Strauss}, {SubbaRao}, {Szalay}, {Thakar}, {Tucker},
  {Vanden Berk}, {Yanny}, {Adelman}, {Anderson}, {Anderson}, {Annis},
  {Bahcall}, {Bakken}, {Bartelmann}, {Bastian}, {Bauer}, {Berman},
  {B{\"o}hringer}, {Boroski}, {Bracker}, {Briegel}, {Briggs}, {Brinkmann},
  {Brunner}, {Carey}, {Carr}, {Chen}, {Christian}, {Colestock}, {Crocker},
  {Csabai}, {Czarapata}, {Dalcanton}, {Davidsen}, {Davis}, {Dehnen},
  {Dodelson}, {Doi}, {Dombeck}, {Donahue}, {Ellman}, {Elms}, {Evans}, {Eyer},
  {Fan}, {Federwitz}, {Friedman}, {Fukugita}, {Gal}, {Gillespie}, {Glazebrook},
  {Gray}, {Grebel}, {Greenawalt}, {Greene}, {Gunn}, {de Haas}, {Haiman},
  {Haldeman}, {Hall}, {Hamabe}, {Hansen}, {Harris}, {Harris}, {Harvanek},
  {Hawley}, {Hayes}, {Heckman}, {Helmi}, {Henden}, {Hogan}, {Hogg}, {Holmgren},
  {Holtzman}, {Huang}, {Hull}, {Ichikawa}, {Ichikawa}, {Johnston}, {Kauffmann},
  {Kim}, {Kimball}, {Kinney}, {Klaene}, {Kleinman}, {Klypin}, {Knapp},
  {Korienek}, {Krolik}, {Kron}, {Krzesi{\'n}ski}, {Lamb}, {Leger},
  {Limmongkol}, {Lindenmeyer}, {Long}, {Loomis}, {Loveday}, {MacKinnon},
  {Mannery}, {Mantsch}, {Margon}, {McGehee}, {McKay}, {McLean}, {Menou},
  {Merelli}, {Mo}, {Monet}, {Nakamura}, {Narayanan}, {Nash}, {Neilsen},
  {Newman}, {Nitta}, {Odenkirchen}, {Okada}, {Okamura}, {Ostriker}, {Owen},
  {Pauls}, {Peoples}, {Peterson}, {Petravick}, {Pope}, {Pordes}, {Postman},
  {Prosapio}, {Quinn}, {Rechenmacher}, {Rivetta}, {Rix}, {Rockosi}, {Rosner},
  {Ruthmansdorfer}, {Sandford}, {Schneider}, {Scranton}, {Sekiguchi}, {Sergey},
  {Sheth}, {Shimasaku}, {Smee}, {Snedden}, {Stebbins}, {Stubbs}, {Szapudi},
  {Szkody}, {Szokoly}, {Tabachnik}, {Tsvetanov}, {Uomoto}, {Vogeley}, {Voges},
  {Waddell}, {Walterbos}, {Wang}, {Watanabe}, {Weinberg}, {White}, {White},
  {Wilhite}, {Wolfe}, {Yasuda}, {York}, {Zehavi}, \& {Zheng}}]{Stoughton:2002}
{Stoughton}, C., {Lupton}, R.~H., {Bernardi}, M., {et~al.} 2002, Astronomical
  Journal, 123, 485

\bibitem[{{Strauss} {et~al.}(2002){Strauss}, {Weinberg}, {Lupton}, {Narayanan},
  {Annis}, {Bernardi}, {Blanton}, {Burles}, {Connolly}, {Dalcanton}, {Doi},
  {Eisenstein}, {Frieman}, {Fukugita}, {Gunn}, {Ivezi{\'c}}, {Kent}, {Kim},
  {Knapp}, {Kron}, {Munn}, {Newberg}, {Nichol}, {Okamura}, {Quinn}, {Richmond},
  {Schlegel}, {Shimasaku}, {SubbaRao}, {Szalay}, {Vanden Berk}, {Vogeley},
  {Yanny}, {Yasuda}, {York}, \& {Zehavi}}]{Strauss:2002}
{Strauss}, M.~A., {Weinberg}, D.~H., {Lupton}, R.~H., {et~al.} 2002,
  Astronomical Journal, 124, 1810

\bibitem[{{SubbaRao} {et~al.}(2002){SubbaRao}, {Frieman}, {Bernardi},
  {Loveday}, {Nichol}, {Castander}, \& {Meiksin}}]{Subbarao:2002}
{SubbaRao}, M., {Frieman}, J., {Bernardi}, M., {et~al.} 2002, in Society of
  Photo-Optical Instrumentation Engineers (SPIE) Conference Series, Vol. 4847,
  Society of Photo-Optical Instrumentation Engineers (SPIE) Conference Series,
  ed. J.-L. {Starck} \& F.~D. {Murtagh}, 452--460

\bibitem[{{Tonry} \& {Davis}(1979)}]{Tonry:1979}
{Tonry}, J. \& {Davis}, M. 1979, Astronomical Journal, 84, 1511

\bibitem[{Yamada(2001)}]{Yamada:2001}
Yamada, I. 2001, in Inherently Parallel Algorithms in Feasibility and
  Optimization and Their Applications, ed. D.~Butnariu, Y.~Censor, \& S.~Reich
  (Elsevier)

\bibitem[{{Zoubian} \& {Kneib}(2013)}]{Zoubian:2013}
{Zoubian}, J. \& {Kneib}, J.-P. 2013, in prep.

\end{thebibliography}
% \label{lastpage}

\end{document}